\documentclass[twoside,twocolumn,9pt]{article}
\usepackage{extsizes}
\usepackage[super,sort&compress,comma]{natbib} 
\usepackage[version=3]{mhchem}
\usepackage[left=1.5cm, right=1.5cm, top=1.785cm, bottom=2.0cm]{geometry}
\usepackage{balance}
\usepackage{mathptmx}
\usepackage{sectsty}
\usepackage{graphicx} 
\usepackage{lastpage}
\usepackage[format=plain,justification=justified,singlelinecheck=false,font={stretch=1.125,small,sf},labelfont=bf,labelsep=space]{caption}
\usepackage{float}
\usepackage{fancyhdr}
\usepackage{fnpos}
\usepackage[english]{babel}
\addto{\captionsenglish}{%
  
}
\usepackage{array}
\usepackage{charter}
\usepackage[T1]{fontenc}
\usepackage[usenames,dvipsnames]{xcolor}
\usepackage{setspace}
\usepackage[compact]{titlesec}
\usepackage{hyperref}

\usepackage{epstopdf}

\usepackage{subcaption}
\usepackage{amsmath}
\usepackage{mathtools}
\usepackage{amssymb}
\usepackage[mathscr]{eucal}

\newcommand{\tr}{\text{tr}}

\newcommand{\nhat}{\hat{\mathbf{n}}}
\newcommand{\mhat}{\hat{\mathbf{m}}}
\newcommand{\lhat}{\hat{\mathbf{l}}}

\newcommand{\Q}{\mathbf{Q}}
\newcommand{\bR}{\mathbf{R}}
\newcommand{\That}{\hat{\mathbf{T}}}
\newcommand{\Omegahat}{\hat{\boldsymbol{\Omega}}}
\newcommand{\bLambda}{\boldsymbol{\Lambda}}

\DeclareMathOperator{\atantwo}{atan2}

\DeclareRobustCommand{\divby}{%
  \mathrel{\vbox{\baselineskip.65ex\lineskiplimit0pt\hbox{.}\hbox{.}\hbox{.}}}%
}

\definecolor{cream}{RGB}{222,217,201}

\begin{document}

\pagestyle{fancy}
\thispagestyle{plain}
\fancypagestyle{plain}{
\renewcommand{\headrulewidth}{0pt}
}

\makeFNbottom
\makeatletter
\renewcommand\LARGE{\@setfontsize\LARGE{15pt}{17}}
\renewcommand\Large{\@setfontsize\Large{12pt}{14}}
\renewcommand\large{\@setfontsize\large{10pt}{12}}
\renewcommand\footnotesize{\@setfontsize\footnotesize{7pt}{10}}
\makeatother

\renewcommand{\thefootnote}{\fnsymbol{footnote}}
\renewcommand\footnoterule{\vspace*{1pt}%
\color{cream}\hrule width 3.5in height 0.4pt \color{black}\vspace*{5pt}} 
\setcounter{secnumdepth}{5}

\makeatletter 
\renewcommand\@biblabel[1]{#1}            
\renewcommand\@makefntext[1]%
{\noindent\makebox[0pt][r]{\@thefnmark\,}#1}
\makeatother 
\renewcommand{\figurename}{\small{Fig.}~}
\sectionfont{\sffamily\Large}
\subsectionfont{\normalsize}
\subsubsectionfont{\bf}
\setstretch{1.125} 
\setlength{\skip\footins}{0.8cm}
\setlength{\footnotesep}{0.25cm}
\setlength{\jot}{10pt}
\titlespacing*{\section}{0pt}{4pt}{4pt}
\titlespacing*{\subsection}{0pt}{15pt}{1pt}

\fancyfoot{}
\fancyfoot[RO]{\footnotesize{\sffamily{1--\pageref{LastPage} ~\textbar  \hspace{2pt}\thepage}}}
\fancyfoot[LE]{\footnotesize{\sffamily{\thepage~\textbar\hspace{3.45cm} 1--\pageref{LastPage}}}}
\fancyhead{}
\renewcommand{\headrulewidth}{0pt} 
\renewcommand{\footrulewidth}{0pt}
\setlength{\arrayrulewidth}{1pt}
\setlength{\columnsep}{6.5mm}
\setlength\bibsep{1pt}

\makeatletter 
\newlength{\figrulesep} 
\setlength{\figrulesep}{0.5\textfloatsep} 

\newcommand{\topfigrule}{\vspace*{-1pt}%
\noindent{\color{cream}\rule[-\figrulesep]{\columnwidth}{1.5pt}} }

\newcommand{\botfigrule}{\vspace*{-2pt}%
\noindent{\color{cream}\rule[\figrulesep]{\columnwidth}{1.5pt}} }

\newcommand{\dblfigrule}{\vspace*{-1pt}%
\noindent{\color{cream}\rule[-\figrulesep]{\textwidth}{1.5pt}} }

\makeatother

\twocolumn[
  \begin{@twocolumnfalse}
\vspace{1em}
\sffamily
\begin{tabular}{m{4.5cm} p{13.5cm} }

& \noindent\LARGE{\textbf{Chiral ground states in a nematic liquid crystal confined to a cylinder with homeotropic anchoring}} \\
\vspace{0.3cm} & \vspace{0.3cm} \\

 & \noindent\large{Lucas Myers$^{\ast}$\textit{$^{a}$} and Jorge Vi\~nals\textit{$^{a}$}} \\

& \noindent\normalsize{
The singular potential method in the $\mathbf{Q}$ tensor order parameter representation is used to determine the ground state configuration of an elastically anisotropic nematic liquid crystal when confined to a cylindrical geometry with homeotropic anchoring. Ground states of broken chiral symmetry are found for sufficiently small values of the twist elastic constant relative to bend and splay constants. For small cylinder radius, twisted configurations, which feature two disclinations lines that wind around the long axis of the cylinder, are generally found to minimize the free energy of the nematic. For larger radii, ground state configurations are (non singular) escaped configuration. Twisted and untwisted escaped configurations are almost degenerate in energy in this region. This near degeneracy is broken when splay-bend contrast is allowed.
} \\

\end{tabular}

 \end{@twocolumnfalse} \vspace{0.6cm}

  ]

\renewcommand*\rmdefault{bch}\normalfont\upshape
\rmfamily
\section*{}
\vspace{-1cm}


\footnotetext{\textit{$^{a}$~School of Physics and Astronomy, University of Minnesota, Minneapolis, MN 55455, USA. E-mail: myers716@umn.edu}}




\section{Introduction}
The ground state of a nematic liquid crystal in a cylindrical geometry, and subject to homeotropic boundary conditions, is investigated numerically within the tensor order parameter representation. A singular potential method is used that allows consideration of defected configurations in three dimensions (containing disclination lines), as well as elastic anisotropy (unequal values of splay, twist, and bend elastic constants).
States of broken chiral symmetry are found to be stable for sufficiently small values of the twist elastic constant. For small radii, twisted configurations, which include two disclinations lines that wind around the long axis of the cylinder, are found to minimize the free energy of the nematic. For large radii, non singular (\lq\lq escaped'') configurations are found to minimize the free energy instead. In this limit, twisted and untwisted configurations are found to be almost degenerate when bend and splay elastic constants are equal.
This degeneracy is lifted when bend-splay anisotropy is allowed, with the twisted state becoming of lower free energy.

Chirality (the absence of mirror symmetry) is a common feature of many soft and living systems \cite{re:ozturk23}, and it is a widely used material property in fields such as catalysis \cite{re:noyori02,re:mackenzie21} and optical sensing \cite{re:brandt17}.
In many systems the molecular units are themselves chiral, a fact that accounts for the appearance of macroscopic handedness. 
However, the appearance of chirality from centrosymmetric molecular units is a more complex phenomenon as it requires the spontaneous breaking of mirror symmetry, and it is often tied to confining effects \cite{re:tortora11,re:jeong14,re:zhang24}. 
Understanding the mechanisms behind spontaneous chiral symmetry breaking is important in the development and application of related technologies in a number of disciplines. 
Our focus here is on the nematic phase of liquid crystals, systems that can be readily controlled and manipulated experimentally, and are well understood theoretically. 
Therefore they offer an excellent platform for the quantitative elucidation of spontaneous chiral symmetry breaking.

Recent experiments involving lyotropic chromonic liquid crystals in a cylindrical capillary have revealed unexpected ground state configurations that break chiral symmetry even though the nematogens themselves are achiral. A lyotropic chromonic liquid crystal is comprised of stacks of disc shaped molecular units that form cylindrical aggregates in solution due to hydrophobic interactions. The stacks are themselves weakly charged.
When the concentration of discs is sufficiently large, and aggregates grow longer, a conventional nematic phase emerges \cite{re:park12,re:kim13,re:collins15,re:zhou17}. For even larger concentrations, the system exhibits a columnar phase, a two dimensional solid. A noteworthy feature of chromonics in their nematic phase is that the twist elastic constant is about one order of magnitude smaller than splay and bend elastic constants, which themselves differ, albeit by a smaller amount \cite{re:zhou17}. The experiments considered planar anchoring on the boundaries (aggregates parallel to the boundary), and revealed an unexpected twisted configuration of the nematic director instead of a ground state configuration with a uniform nematic director field that is everywhere parallel to the long axis of the capillary \cite{nayani2015spontaneous}.
Similar phenomenology has been observed by others \cite{re:davidson15}, including in rectangular capillaries \cite{re:fu17} and in cylindrical shells \cite{re:javadi18}. Closer to our analysis below, chiral configurations have also been observed in capillaries with homeotropic anchoring on boundaries (aggregates perpendicular to the boundary) \cite{jeong_chiral_2015}, and in nematic micellar systems \cite{dietrich_chiral_2017}, also with homeotropic anchoring. Just like lyotropic chromonics, micellar systems also feature a very small twist elastic constant relative to splay and bend \cite{re:dietrich20}.

When the chromonic in the nematic phase is described by a director field, with energies of distortion given by the classical Frank free energy, the experimental observation of chiral phases under planar anchoring led to the conclusion that chromonics violate one classical Ericksen inequality \cite{ericksen_1966}. This violation was associated with the anomalously small value of the twist elastic constant determined \cite{nayani2015spontaneous}. Since then, however, it has been shown theoretically that twisted ground states may be thermodynamically stable even for elastic constants which violate the weak form of the inequalities (for stability of a uniform, infinite, system), as long as the system is confined to a particular geometry \cite{long_violation_2023}. Corroborating local stability results have also been given \cite{re:paparini22}.

\begin{figure}
    \centering
    \hfill
    \begin{subfigure}{0.5\columnwidth}
        \includegraphics[width=\textwidth]{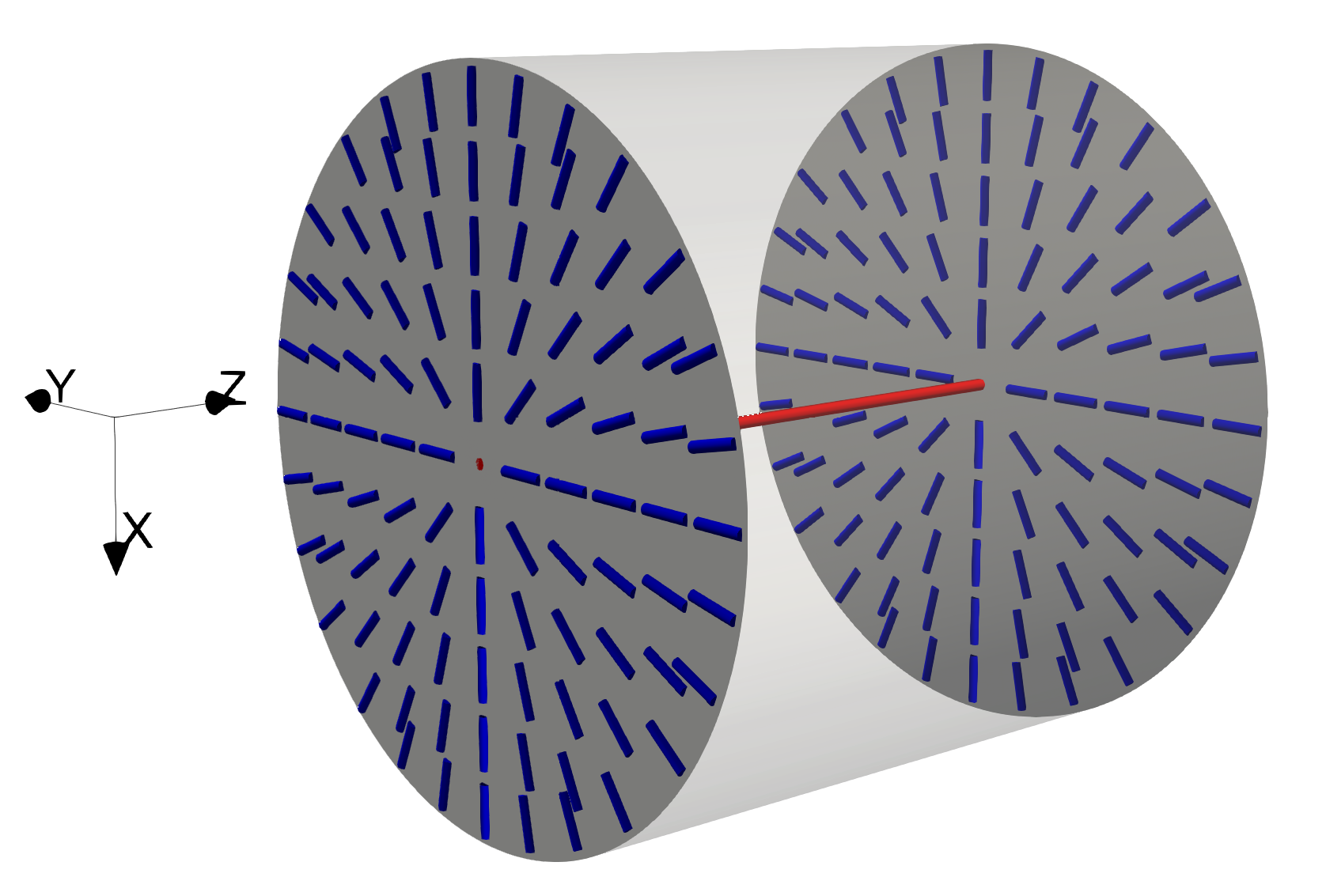}
        \caption{}
        \label{fig:PR-diagram}
    \end{subfigure}
    \hfill
    \begin{subfigure}{0.4\columnwidth}
        \includegraphics[width=\textwidth]{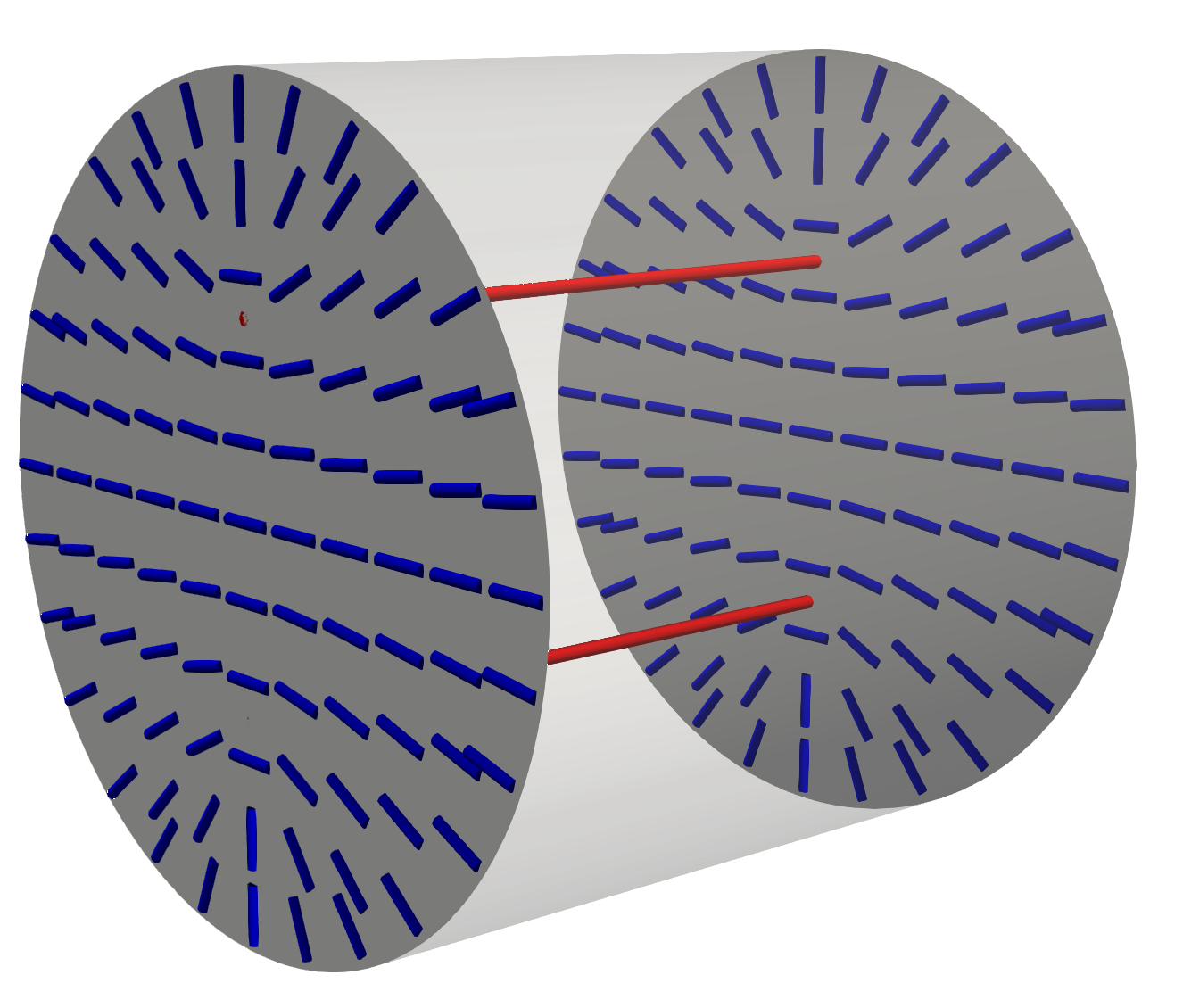}
        \caption{}
        \label{fig:PP-diagram}
    \end{subfigure}

    \hfill
    \begin{subfigure}{0.45\columnwidth}
        \includegraphics[width=\textwidth]{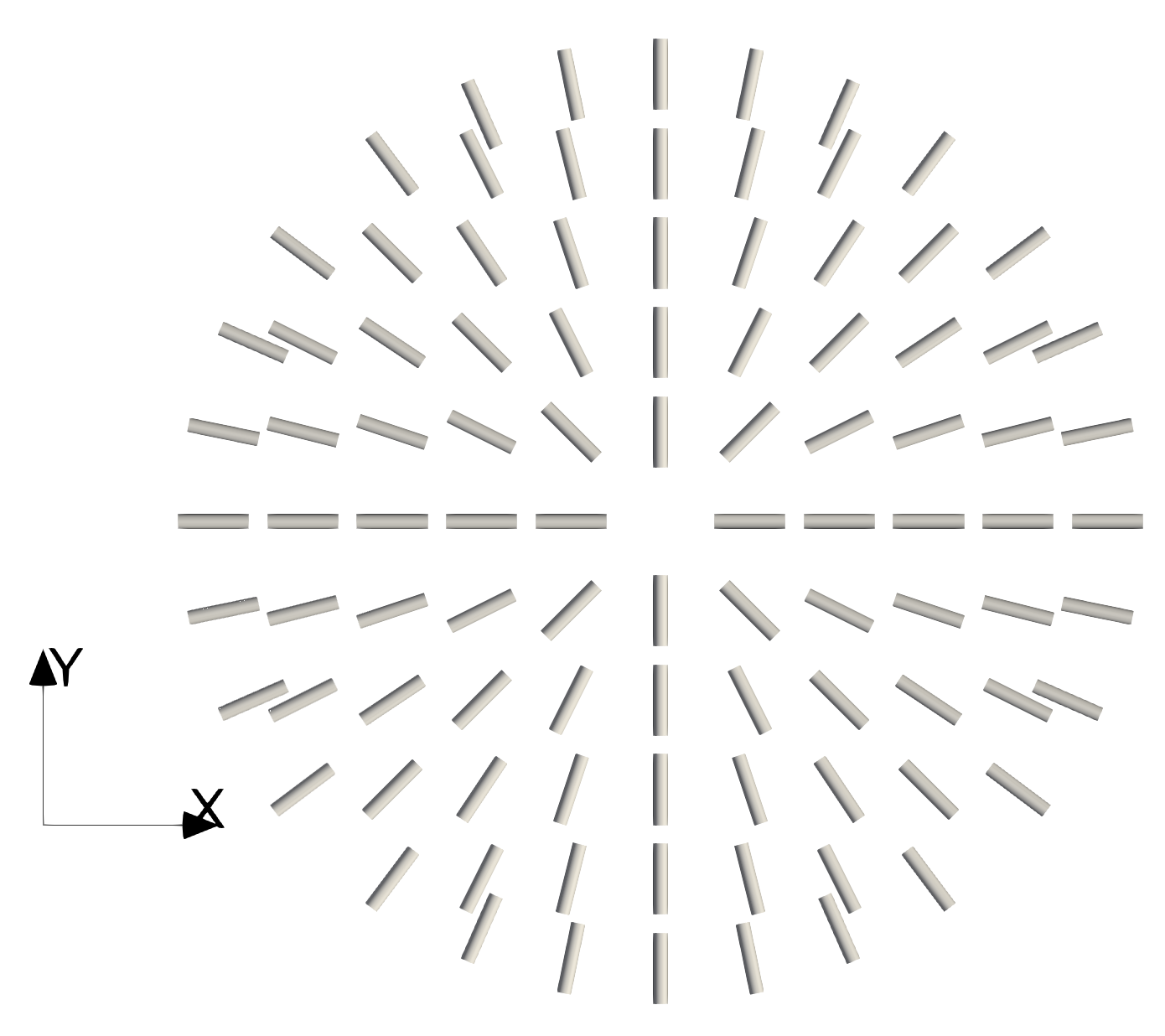}
        \caption{}
        \label{fig:PR-x-y}
    \end{subfigure}
    \hfill
    \begin{subfigure}{0.4\columnwidth}
        \includegraphics[width=\textwidth]{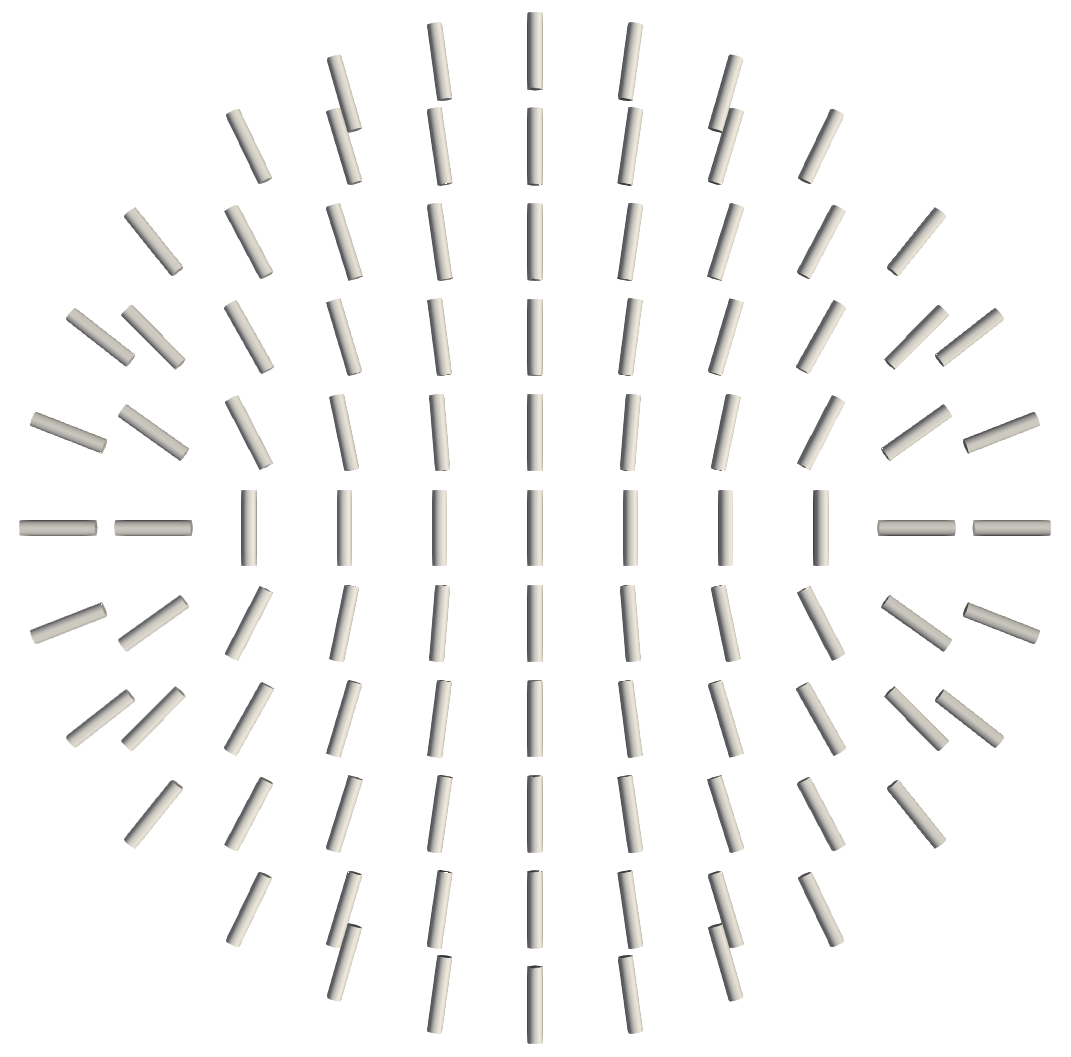}
        \caption{}
        \label{fig:PP-x-y}
    \end{subfigure}

    \caption{Diagrams of PR (a, c) and PP (b, d) configurations.
    In (a) and (b) disclination lines are depicted in red while director orientations are depicted in blue for representative $x$-$y$ cross-sections.
    }
\end{figure}

\begin{figure}
    \centering
    \hfill
    \begin{subfigure}{0.45\columnwidth}
        \includegraphics[width=\textwidth]{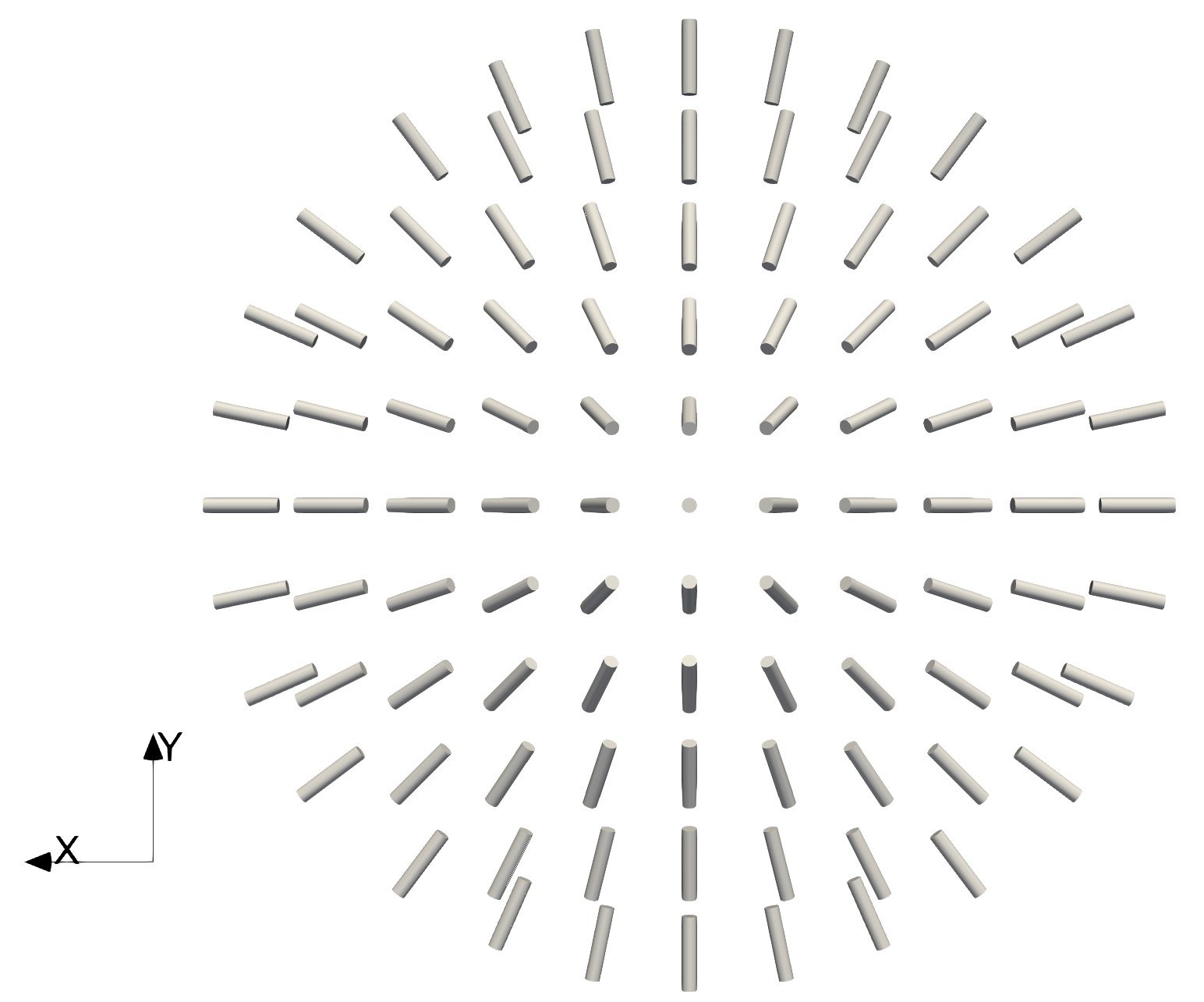}
        \caption{}
        \label{fig:ER-x-y}
    \end{subfigure}
    \hfill
    \begin{subfigure}{0.40\columnwidth}
        \includegraphics[width=\textwidth]{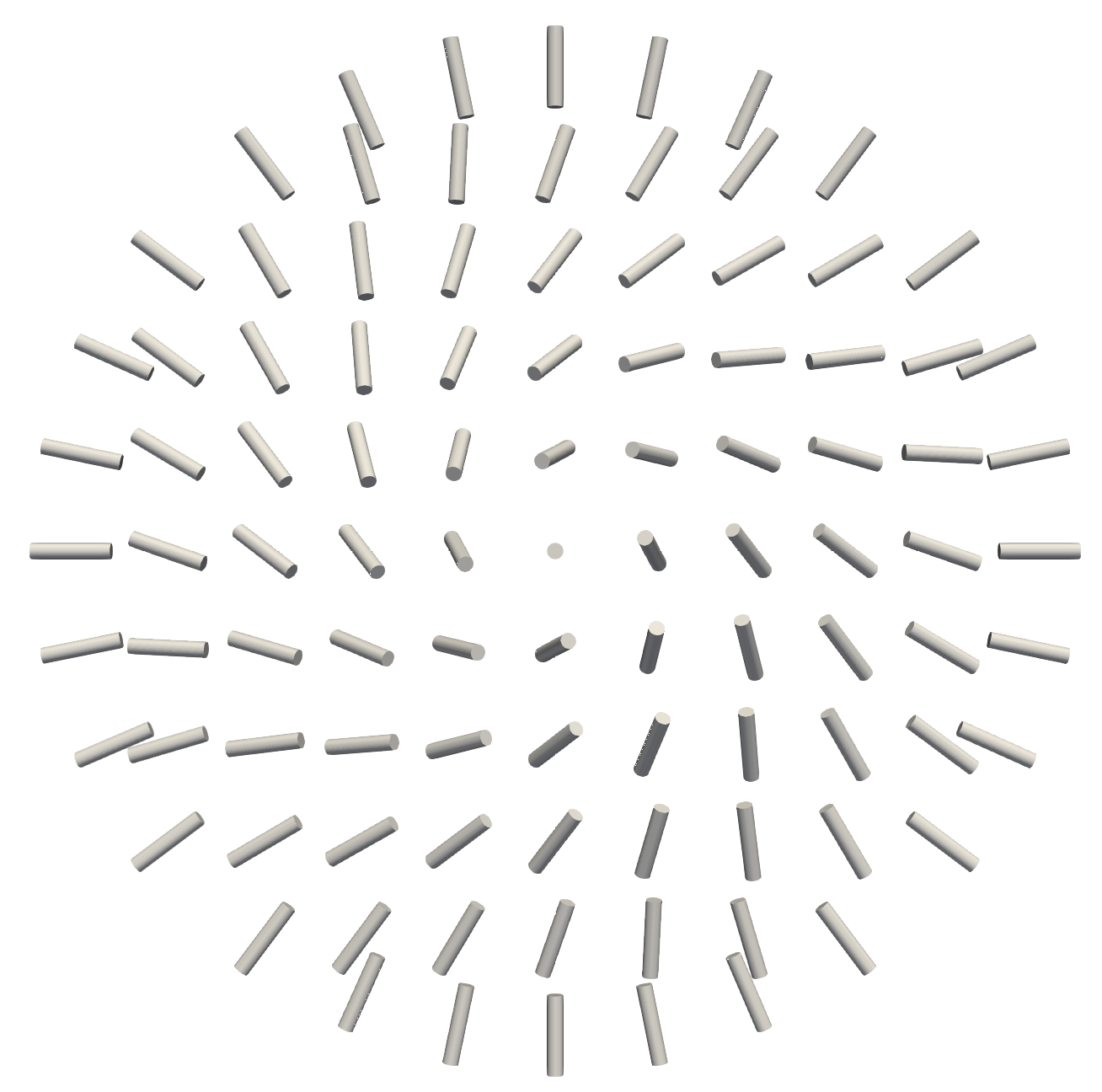}
        \caption{}
        \label{fig:TER-x-y}
    \end{subfigure}

    \hfill
    \begin{subfigure}{0.45\columnwidth}
        \includegraphics[width=\textwidth]{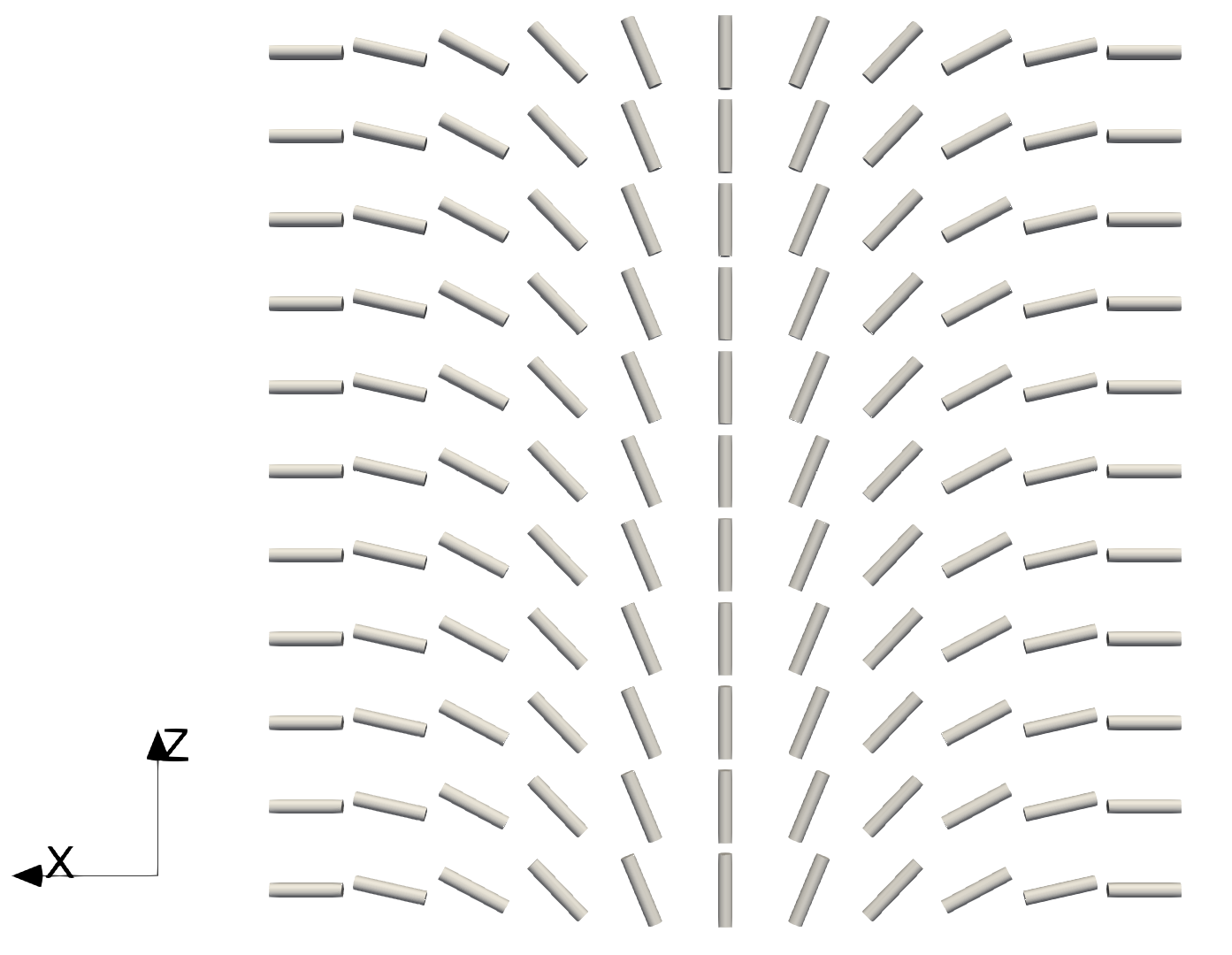}
        \caption{}
        \label{fig:ER-x-z}
    \end{subfigure}
    \hfill
    \begin{subfigure}{0.37\columnwidth}
        \includegraphics[width=\textwidth]{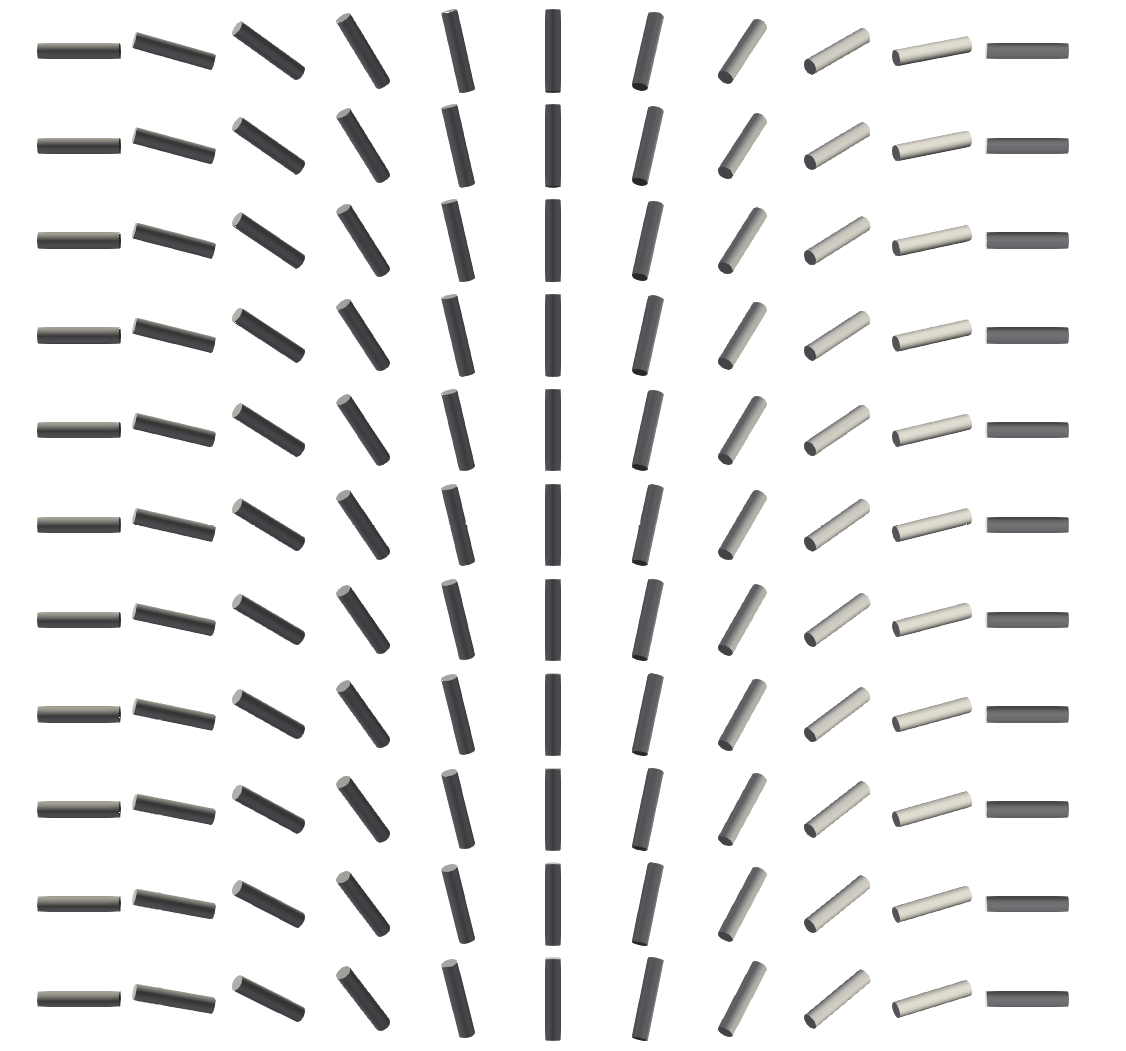}
        \caption{}
        \label{fig:TER-x-z}
    \end{subfigure}

    \caption{Diagrams of ER (a, c) and TER (b, d) configurations. 
      The director escapes in the $-z$ direction for both configurations, and the TER configuration twists along the escape direction.}
    \label{fig:ER}
\end{figure}

The nematic ground state in a cylinder under homeotropic anchoring has already been studied numerically in the isotropic (one constant) limit \cite{re:yan02,shams_theoretical_2014}. For the narrowest capillaries, the so called polar radial configuration (PR) (Figs. \ref{fig:PR-diagram}, \ref{fig:PR-x-y}) was observed with a single +1 disclination line along the cylinder axis. This configuration is not topologically stable, and hence it is expected to decay (\lq\lq escape through the third dimension\rq\rq) into what is known as an escaped radial (ER) configuration (Figs. \ref{fig:ER-x-y}, \ref{fig:ER-x-z}). Nevertheless, when the radius of the capillary is increased, a stable polar planar (PP) configuration (Figs. \ref{fig:PP-diagram}, \ref{fig:PP-x-y}) with two +1/2 parallel disclinations along the long direction of the cylinder were found instead. With further radius increases, the ground state observed in the computations is the ER configuration \cite{shams_theoretical_2014}.

\begin{figure*}
    \centering
    \hfill
    \begin{subfigure}{0.55\textwidth}
        \includegraphics[width=\textwidth]{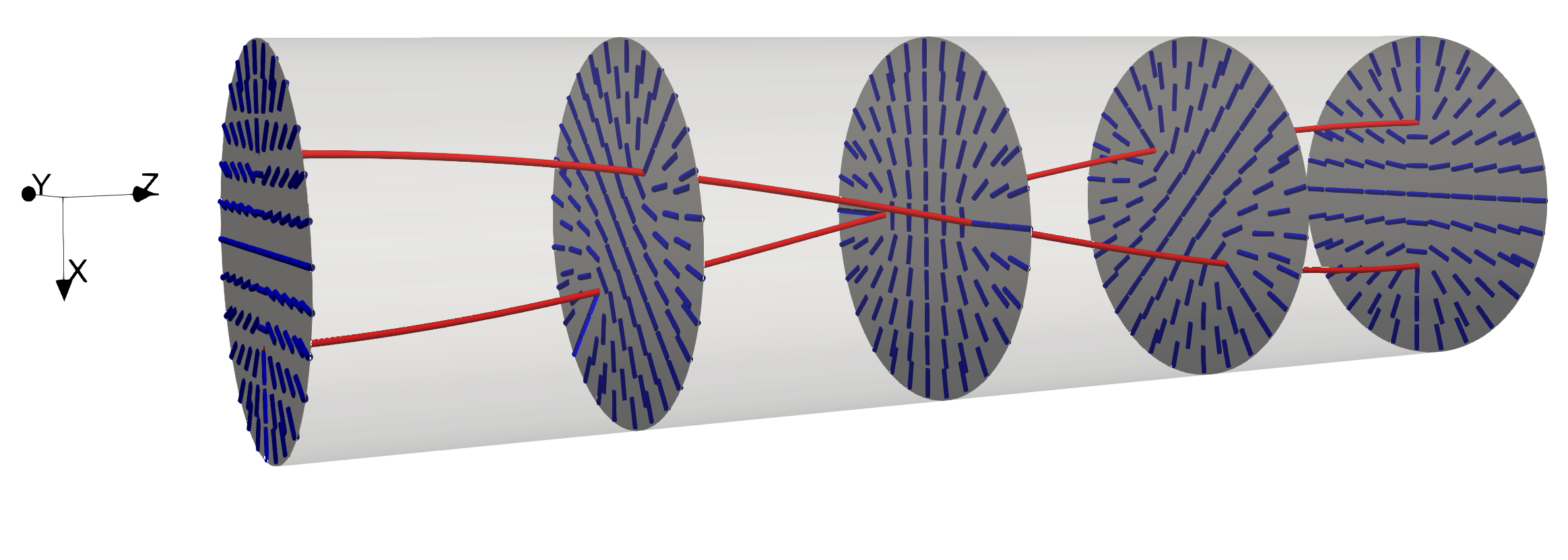}
        \caption{}
        \label{fig:TP-diagram}
    \end{subfigure}
    \hfill
    \begin{subfigure}{0.40\textwidth}
        \includegraphics[width=\textwidth]{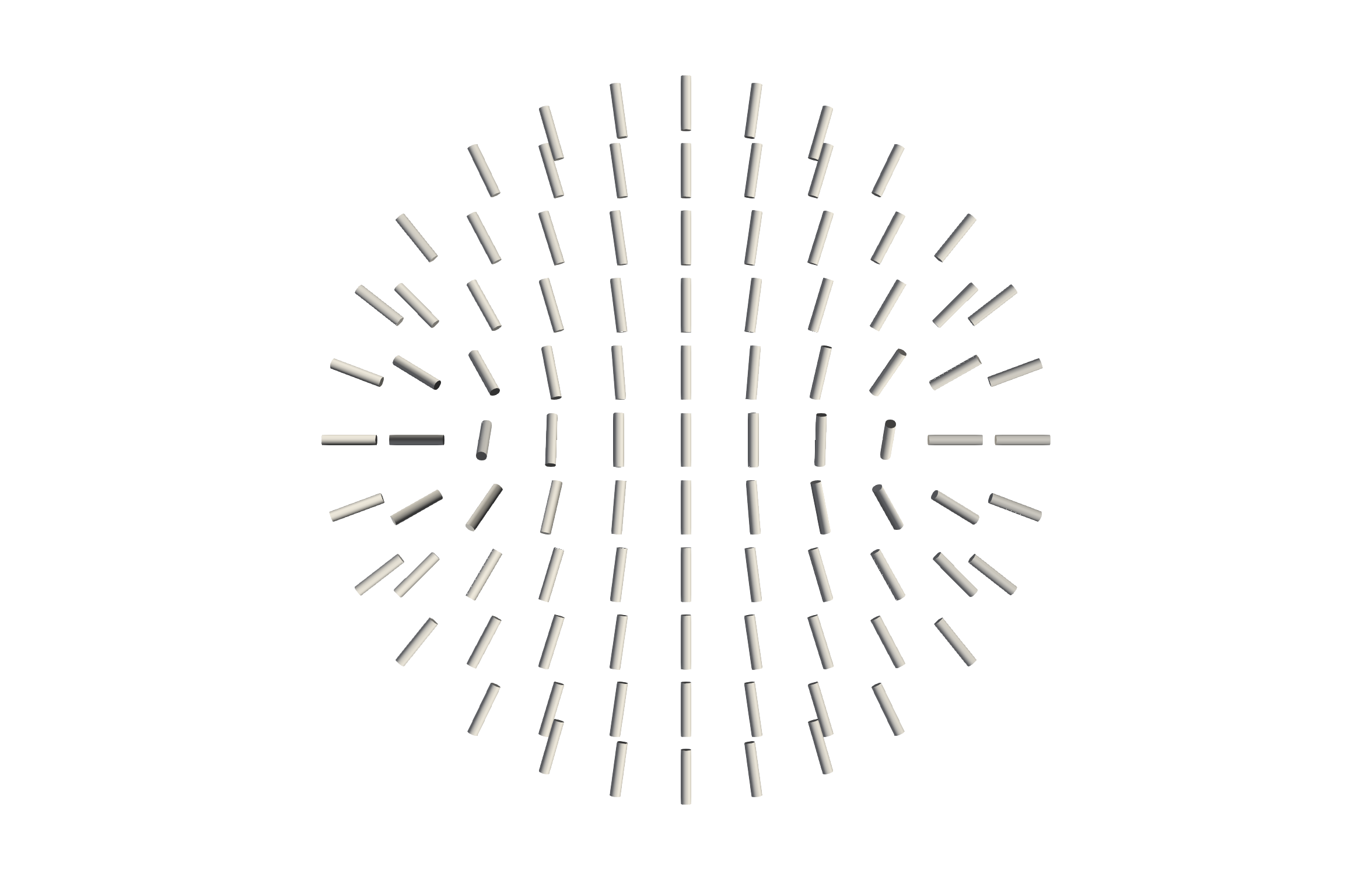}
        \caption{}
        \label{fig:TP-x-y}
    \end{subfigure}

    \caption{Diagram of TPP configuration (a) and cross-section of TP-configuration (b).
    In both the TPP and TP configurations the disclinations form a double-helix structure, as in (a).
    Each cross-section of the TPP configuration is a cross-section of a PP configuration rotated about the cylindrical axis.
    In the TP configuration, the director is rotated near each disclination about the axis which connects the two disclinations.
    This rotation depends on the chirality of the double-helix structure, and tends to align the director at the disclination center to be tangent to the disclination line, as is explained in Section \ref{sec:TP}.}
    \label{fig:twisted-configurations}
\end{figure*}

We extend these calculations below by allowing elastic anisotropy of the nematic, and non planar configurations. In particular, we address the case of small twist elastic constant relative to bend and splay, and the appearance of twisted configurations. In this case, the experimental phenomenology regarding spontaneous chiral symmetry breaking is quite complex. It has been found experimentally that an ER configuration may spontaneously break chiral symmetry to become twisted -- a twisted escaped radial (TER) configuration (Figs. \ref{fig:TER-x-y}, \ref{fig:TER-x-z}) -- which is argued to decay further into a configuration featuring two disclinations forming a double-helix along the capillary \cite{jeong_chiral_2015}.
It was speculated that this configuration consists of two +1/2 disclinations in which the director remains in plane, and was dubbed a twisted polar planar (TPP) configuration (Figs. \ref{fig:PP-x-y}, \ref{fig:TP-diagram}).
Nevertheless, further experiments in nematic micellar systems, while confirming the existence of the double helix configuration, concluded that the director escapes out of the plane near the disclination centers, leading to the so called twisted polar (TP) configuration (Fig. \ref{fig:twisted-configurations}) \cite{dietrich_chiral_2017}.

We show below that all chiral configurations include director twist near the defect cores or near the cylinder center in the escaped cases, and that this twist is necessary for symmetry breaking. Free energy that would be otherwise contained in splay or bend modes is transferred to twist for sufficiently large elastic constant contrast. Our numerical analysis is based on a tensor order parameter representation of the nematic, which is free of the limitations associated with the Frank free energy model at dealing with disclinations in three dimensions. In addition, we use a singular potential theory \cite{re:ball10,re:schimming21} to consistently accommodate elastic anisotropy. This allows us to compare the free energies of the twisted and corresponding untwisted states, and to obtain a stability diagram of each of the nematic states which may exist in cylindrical capillaries with homeotropic anchoring.

\section{Singular potential method calculation of the free energy}

In the director representation, the average local orientation of nematic molecules is described by the director field, a unit vector $\nhat (\mathbf x)$. The Frank free energy penalizes distortions away from a uniform ground state and consists of all scalar terms up to second order which are composed of gradients of $\nhat$, and which respect nematic symmetry $\nhat \to -\nhat$. It reads:
\begin{equation}
    F_n \left(\nhat, \nabla \nhat\right)
    =
    \begin{multlined}[t]
        \int_\Omega
        \biggl[
            \frac12 K_1 \left( \nabla \cdot \nhat \right)^2
            + \frac12 K_2 \left[ \nhat \cdot \left( \nabla \times \nhat \right) \right]^2 \\
            + \frac12 K_3 \left| \nhat \times \left( \nabla \times \nhat \right) \right|^2
            + \frac12 K_{24} \nabla \cdot \left[ \left( \nhat \cdot \nabla \right) \nhat - \nhat \left( \nabla \cdot \nhat \right) \right]
        \biggr] dV
    \end{multlined}
\end{equation}
with $K_1$, $K_2$, $K_3$, and $K_{24}$ elastic constants associated with splay, twist, bend, and saddle splay distortion modes respectively \cite{re:selinger16}. This energy diverges near the core of defects as the gradients of the order parameter become arbitrarily large. Cut-off lengths can be introduced to render the energy finite, but this is only well understood for straight disclination lines in an elastically isotropic and uniaxial nematic. In practice, the region surrounding the core is biaxial, and can display significant anisotropy \cite{re:zhou17,re:schimming20}. An alternative representation which circumvents the need to specifically treat director field singularities in the director is the $Q$ tensor model. Nematic order is described is by a $3\times 3$ tensor order parameter field $\Q$ giving the magnitude and direction of local orientational order. The tensor $\Q$ is defined in terms of an equilibrium probability distribution function $\rho \left( \mathbf p \right)$ of nematogen direction $\mathbf p$,
\begin{equation} \label{eq:Q-tensor-definition}
    \mathbf Q
    =
    \int_{S^2}
    \left( \mathbf p \otimes \mathbf p - \frac13 \mathbf I \right) \rho \left( \mathbf p \right) d \sigma
\end{equation}
The domain of integration is the unit two-sphere $S^2$ with surface measure $d\sigma$ because molecular orientation is described by a unit vector $\mathbf p$. Note that $\rho(\mathbf p) = \rho(-\mathbf p)$ due to nematic symmetry, so that the $Q$-tensor is traceless and symmetric. $\Q$ may be diagonalized with real eigenvalues, $\lambda_1 \geq \lambda_2 \geq \lambda_3$, and corresponding orthonormal eigenvectors $\nhat$, $\mhat$, and $\lhat$.  The well known scalar uniaxial order parameter may be defined from the eigenvalues as $S = \frac32 \lambda_1$. The biaxial order parameter is given by $P = \frac12 \lambda_1 + \lambda_2$. Disclination lines are now given as the line in which two positive eigenvalues become degenerate.

The free energy of the nematic can be written in terms of $\Q$, and it often comprises two terms: a bulk free energy of the Landau-de Gennes form \cite{re:selinger16}, and an elastic contribution that depends on spatial gradients of $\Q$ (see also Eq. \eqref{eq:fe_elastic}) \cite{re:schimming21}. To lowest (second) order in a gradient expansion of the free energy as a function of $\Q$, the elastic energy may exhibit twist elastic anisotropy ($L_2 \neq 0$), but no splay-bend anisotropy (Eqs. \eqref{eq:KL_mapping} with $L_{3} = 0$). Third order terms in $\Q$ break this degeneracy, but the free energy at this order becomes unbounded for all values of its parameters \cite{re:ball10,re:bauman16}. A stable free energy implies consideration of terms at least of fourth order in $\Q$. However, there are 22 possible terms allowed by symmetry up to fourth order \cite{re:longa87}, thus making the theory intractable for anisotropic systems. Such a lack of stability can be traced back to the fact that the Landau-de Gennes theory, as formulated, does not constrain the eigenvalues of $\Q$ to remain within their physically admissible range \cite{re:ball10}. From the definition, Eq. \eqref{eq:Q-tensor-definition}, this range is $  -S/3 \le \lambda_{i} \le 2S/3  $. The singular potential method is devised so as to enforce this constraint on the eigenvalues of $\Q$.

The singular potential method considers a bulk free energy $F_b[\Q] = E[\Q] - T \Delta \mathscr{S}[\Q]$ where the internal energy is taken to be of the Maier-Saupe form $E[\Q] = -\kappa \int_{\Omega} \tr \left[ \Q \left( \mathbf x \right)^2\right] dV$, with $\kappa$ a positive constant which characterizes alignment strength \cite{re:selinger16}. A microscopic definition of the entropy difference between a nematic and an isotropic configuration is introduced as
\begin{equation} \label{eq:entropy-definition}
    \Delta \mathscr S
    =
    -n k_B \int_{\Omega} \int_{S^2} \rho(\mathbf p, \mathbf x) \ln \left[ 4\pi \rho(\mathbf p, \mathbf x) \right] d\sigma \, dV
\end{equation}
with $n$ the number density of nematogens, $k_B$ Boltzmann's constant, $\Omega$ the spatial domain, and $T$ the uniform temperature. The entropy is now maximized over all microscopic configurations that yield a specified tensor $\Q$ according to Eq. \eqref{eq:Q-tensor-definition}. If $\bLambda$ is a tensor of Lagrange multipliers (also traceless and symmetric), the distribution that maximizes the entropy is
\begin{equation} \label{eq:molecular-pdf}
    \rho(\mathbf p)
    =
    \frac{\exp \left( \mathbf p^T \bLambda \mathbf p \right)}{Z[\bLambda]}, \quad\quad
    Z[\bLambda]
    =
    \int_{S^2} \exp\left( \mathbf p^T \bLambda \mathbf p\right) d\sigma
  \end{equation}
with partition function $Z[\bLambda]$. By substituting Eq. \eqref{eq:molecular-pdf} into \eqref{eq:Q-tensor-definition} we may relate $\bLambda$ to $\Q$ through the self consistency condition,
\begin{equation} \label{eq:implicit-singular-potential-equation}
    \Q
    =
    \frac{\partial \ln Z}{\partial \bLambda} - \frac13 \mathbf I
\end{equation}
By substituting Eq. \eqref{eq:molecular-pdf} into Eq. \eqref{eq:entropy-definition}, the constrained entropy may be written in terms of both $\Q$ and $\bLambda$,
\begin{equation}
    \Delta \mathscr S
    =
    -n k_B \int_{\Omega} \left[
        \ln 4\pi
        - \ln Z[\Q]
        + \bLambda[\Q] : \left(
            \Q + \frac13 I
        \right)
    \right]
    dV
\end{equation}
with $:$ denoting a double index contraction. Both tensors are not independent, but related through Eq. \eqref{eq:implicit-singular-potential-equation}. The partition function $Z[\bLambda]$ needs to be evaluated numerically, adding to the complexity of the method \cite{re:schimming21}.

For the elastic free energy, we consider here only one term of third order in $\Q$ to allow for bend-splay anisotropy,
\begin{equation}
    F_\text{el}
    =
    \int_{\Omega}
   \left[
        L_1 \left| \nabla \Q \right|^2
        + L_2 \left| \nabla \cdot \Q \right|^2
        + L_3 \left( \nabla \Q \right) \divby \left[ \left( \Q \cdot \nabla \right) \Q \right]
    \right]
    dV
    \label{eq:fe_elastic}
\end{equation}
with $\divby$ a triple index contraction from inner indices to outer indices, and $L_i$ are the elastic constants.
In index notation, the energy reads,
\begin{equation}
    F_\text{el}[\Q, \nabla \Q]
    =
    \int_{\Omega}
    \left[
        L_1 \left( \partial_k Q_{ij} \right)^2
        + L_2 \left( \partial_j Q_{ij} \right)^2
        + L_3 Q_{lk} \left( \partial_l Q_{ij} \right) \left( \partial_k Q_{ij} \right)
    \right]
\end{equation}
For a uniaxial, constant $S$ configuration, a correspondence may be drawn between the Frank energy and Landau-de Gennes elastic energy coefficients as follows,
\begin{equation}
\begin{split}
    K_1
    &=
    4 L_1 S^2 + 2 L_2 S^2 - \frac43 L_3 S^3 , \quad\quad K_2 =  4 L_1 S^2 - \frac43 L_3 S^3 \\
    K_3
    &=
   4 L_1 S^2 + 2 L_2 S^2 + \frac83 L_3 S^3,  \quad\quad
    K_{24} =
    4 L_1 S^2 - \frac43 L_3 S^3
  \end{split}
  \label{eq:KL_mapping}
\end{equation}
The full free energy is then given by $F[\Q, \nabla \Q] = F_b[\Q] + F_\text{el}[\Q, \nabla \Q]$.

In order to find minimizers of $F[\Q, \nabla \Q]$, we will solve a rotational diffusion equation in time $t$,
\begin{equation} \label{eq:Q-tensor-eom}
    \frac{\partial \Q}{\partial t}
    =
    -\gamma \frac{\delta F}{\delta \Q}
\end{equation}
subject to homeotropic boundary conditions until a steady state is reached. The constant $\gamma$ is a rotational diffusion coefficient. Dimensionless variables are introduced according to $\overline x = x / \xi, \: \overline t = t / \tau, \: \overline \kappa = \frac{2 \kappa}{n k_B T}, \: \overline L_2 = \frac{L_2}{L_1}, \: \overline L_3 = \frac{L_3}{L_1}$ where length and time scales given by,  $\xi = \sqrt{\frac{2 L_1}{n k_B T}}, \: \tau = \frac{1}{\gamma n k_B T}$. We drop the overlines for brevity, and all quantities are given in these dimensionless length and time scales.

For configurations which are uniform along the cylindrical axis (PR, PP, ER, TER), Eq. \eqref{eq:Q-tensor-eom} is solved on a two-dimensional disc with the field fixed along the boundary to be uniaxial and constant $S = S_0$, with director perpendicular to the boundary. 
Here $S_0$ is the equilibrium value of $S$ for a uniform configuration as determined by $\kappa$. 
Note that solving Eq. \eqref{eq:Q-tensor-eom} for a non planar configuration on a two-dimensional disc is equivalent to solving on an infinite cylinder under the condition that the configuration be uniform along the cylindrical axis imposed. 
For the TP and TPP configurations, Eq. \eqref{eq:Q-tensor-eom} is solved on a three-dimensional cylinder with field fixed along the curved boundary to be uniaxial and constant $S = S_0$, with director perpendicular to the curved boundary. 
These configurations are initialized with some fixed wavenumber $\omega$ which determines the pitch. 
For reasons discussed in Section \ref{sec:TP}, the length of the cylinder is chosen to be half of the pitch of the initialized configuration, and periodic boundary conditions are imposed on the cylindrical caps.

In order to numerically solve Eq. \eqref{eq:Q-tensor-eom} we discretize it in time with a Crank-Nicolson method, and in space by using a finite element method with a quadrilateral mesh, and first order Lagrangian elements. The resulting equation is nonlinear in $\Q$, and we use a Newton-Rhapson method to solve for $\Q$ at each time step. The singular potential $\bLambda$ is not analytically tractable, and Eq. \eqref{eq:implicit-singular-potential-equation} is evaluated numerically at each point in space by using a Newton-Rhapson iteration. The integrals over the sphere $S^{2}$ are evaluated with a Lebedev quadrature scheme. Configurations are iterated in time until energy is approximately stationary.  The numerical method is implemented using the deal.II finite element framework \cite{re:heltai21, re:ardnt21}. For more details on the numerical method and the code used in this work, see \cite{myers_2024, re:myers25}.

\section{Thin capillaries: double coiled chiral configurations} \label{sec:TP}

In a study of a lyotropic chromonic (Sunset Yellow) confined to a capillary with homeotropic anchoring \cite{jeong_chiral_2015}, a defect free (escaped) but twisted configuration (TER) has been reported to decay into a configuration that features two line disclinations along the long axis of the cylinder that coil around each other forming a double helix structure. 
One possibility is that this configuration may be the so called twisted planar polar (TPP) configuration in which the disclination lines coil into a double helix, but the director remains confined to the plane perpendicular to the cylindrical axis. 
A similar double helix structure is found in nematic micelles \cite{dietrich_chiral_2017}, except that an analysis with crossed polarizers reveals that the director does not remain planar near the disclination cores. 
This configuration, with out of plane director, was named twisted polar (TP).
We wish to address two issues in this section: firstly, whether the coiled disclination configurations are a true ground state as compared to the straight, parallel disclination configurations, and if so under what conditions.
Secondly, if such coiled configurations are a true ground state, whether this implies that the emergence of chirality is accompanied by director twist near the core. 
We find that for large $L_2$, the free energy is minimized in the coiled configuration with some nonzero wavenumber $\omega$.
Additionally, we find that the ground state that minimizes the free energy in the case of a double helix configuration shows, in fact, director twist near the cores. 
The latter is the mechanism that enables the macroscopic coiling displayed by the configuration in the capillary.

That a double helix configuration must also exhibit director twist can be argued directly in the director representation. Consider the Frank elastic energy of a configuration in which the director remains in plane, and for which $K_1 = K_3 = K \neq K_2$. The choice of the $K_{24}$ saddle-splay elastic constant is arbitrary because the saddle-splay elastic term is manifestly zero for in-plane director configurations. Let $\theta$ be the angle that the director makes in plane with respect to one of the planar axes, and define an elastic twist anisotropy constant $\zeta = \frac{K - K_2}{K + K_2}$. Note that $|\zeta| \leq 1$. The Frank elastic energy reduces to,
\begin{equation}
    F_\text{planar}
    =
    \int_{\Omega} 
    \left(
    \left(1 + \zeta \right) 
    \left[ 
        \left( \partial_x \theta \right)^2 + \left( \partial_y \theta \right)^2
    \right]
    +
    \left( 1 - \zeta \right) \left( \partial_z \theta \right)^2
    \right)
    dV
\end{equation}
The term in square brackets is minimized by the solution to Laplace's equation in two dimensions, and the $z$-derivative term is always non negative. Thus, the total minimizer is a solution to Laplace's equation in two dimensions, and uniform in the $z$-direction. For a configuration with two $+1/2$ disclinations, this indicates that the double-helix structure is energetically unfavorable as compared to the straight, parallel disclination structure, especially considering the increased length of the disclination lines in the former case, which would tend to increase overall configuration energy. This argument indicates that any energetically favorable coiled double helix configuration must be out of plane in some region.

In order to elucidate the spatial structure of such a configuration, fully three dimensional configurations in a cylinder with uniaxial homeotropic boundary conditions are sought that minimize the singular potential free energy in the $\Q$ tensor representation. Periodic boundary conditions are imposed on the cylinder caps to minimize edge effects.
The choice of initial condition in the free energy minimization of Eq. \eqref{eq:Q-tensor-eom} is of great importance because there exist families of almost degenerate configurations (as the pitch of the disclination coiling is continuously changed), but also configurations of different symmetry that have almost the same energy. In the study of this section, we initialize the configuration of the system as a double helix with constant wavenumber $\omega$,  and in plane director perpendicular to the cylindrical axis. Explicitly, a two dimensional $\Q$ tensor configuration is defined,
\begin{equation} \label{eq:Q-2D-definition}
    \Q_{2D} \left(x, y\right)
    =
    q_1 \left( \nhat \otimes \nhat \right)
    + q_2 \left( \mhat \otimes \mhat \right)
    - \left( q_1 + q_2 \right) \left( \lhat \otimes \lhat \right)
\end{equation}
with the following quantities defining a two dimensional configuration including two disclinations on the cross sectional plane,
\begin{align}
    q_1 \left(x, y\right) \label{eq:q1-definition}
    &=
    \left( q_\text{max} - q_\text{min} \right) \left( -1 + \tanh \frac{r_1}{r_0} + \tanh \frac{r_2}{r_0} \right) + q_\text{min} \\
    q_2 \left(x, y \right) \label{eq:q2-definition}
    &=
    \left( q_\text{max} - q_\text{min} \right) \left( 2 - \tanh \frac{r_1}{r_0} - \tanh \frac{r_2}{r_0} \right) + q_\text{min} \\
    \theta \left(x, y\right)
    &=
    \frac12 \left( \varphi_1 +  \varphi_2 \right) \\
    \nhat \left(x, y\right)
    &=
    \begin{bmatrix}
        0 &\cos\theta &\sin\theta
    \end{bmatrix}^T \\
    \mhat \left(x, y \right)
    &=
    \begin{bmatrix}
        0 &-\sin\theta &\cos\theta
    \end{bmatrix}^T
\end{align}
Here $q_\text{max}$ and $q_\text{min}$ are the maximum and minimum values of the largest eigenvalue $\lambda_1 = \frac23 S$ of $\Q$, and are chosen to be $q_\text{max} = \frac23 S_0$, $q_\text{min} = \frac14 q_\text{max}$ with $S_0$ the equilibrium value of $S$ for a uniform configuration, with $\kappa = 8.0$. 
The pairs $(r_1, \varphi_1)$, $(r_2, \varphi_2)$ are polar coordinates centered on each of the disclinations located at $(x_1, y_1)$ and $(x_2, y_2)$ respectively, 
\begin{equation}
    r_i
    =
    \sqrt{\left(x - x_i\right)^2 + \left(y - y_i\right)^2}, \:\:\:
    \varphi_i
    =
    \atantwo \left( y - y_i, x - x_i\right) .
\end{equation}
The unit vector $\lhat$ forms a right-handed orthonormal basis along with $\nhat$ and $\mhat$.
The specific forms of Eqs. \eqref{eq:q1-definition} and \eqref{eq:q2-definition} are chosen so that $\Q$ is continuous through the defect points, and that its eigenvalue profile around each disclination approximates that of a relaxed, isolated, disclination \cite{long_geometry_2021}. The full three dimensional initial condition is obtained by rotating the field about the cylindrical axis,
\begin{equation}
    \Q \left(x, y, z \right)
    =
    \bR \left(\omega z\right) \Q_{2D} \left(x', y' \right) \bR^T \left( \omega z \right)
\end{equation}
with $x', y'$ rotated Cartesian coordinates,
\begin{equation}
    \begin{bmatrix}
        x' \\ y' \\ z'
    \end{bmatrix}
    =
    R^T \left( \omega z \right)
    \begin{bmatrix}
        x \\ y \\ z
    \end{bmatrix}
\end{equation}
The rotation matrix that is used is,
\begin{equation} \label{eq:rotation-matrix-definition}
    R\left(\omega z \right)
    =
    \begin{bmatrix}
        \cos \left( \omega z \right) &-\sin \left( \omega z \right) &0 \\
        \sin \left( \omega z \right) &\cos \left(\omega z \right) &0 \\
        0 &0 &1
    \end{bmatrix}
\end{equation}

\begin{figure}
    \centering
    \begin{subfigure}{\columnwidth}
        \includegraphics[width=\textwidth]{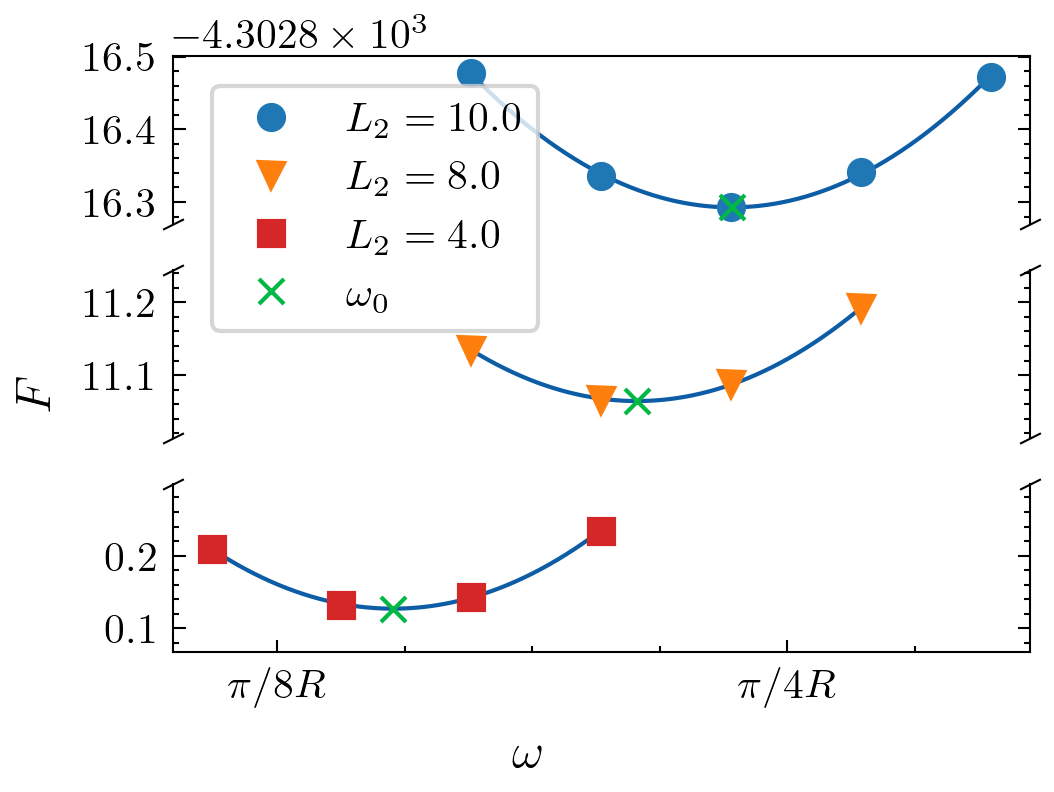}
        \caption{}
        \label{fig:all-L2-plot}
    \end{subfigure}
    \begin{subfigure}{\columnwidth}
        \includegraphics[width=\textwidth]{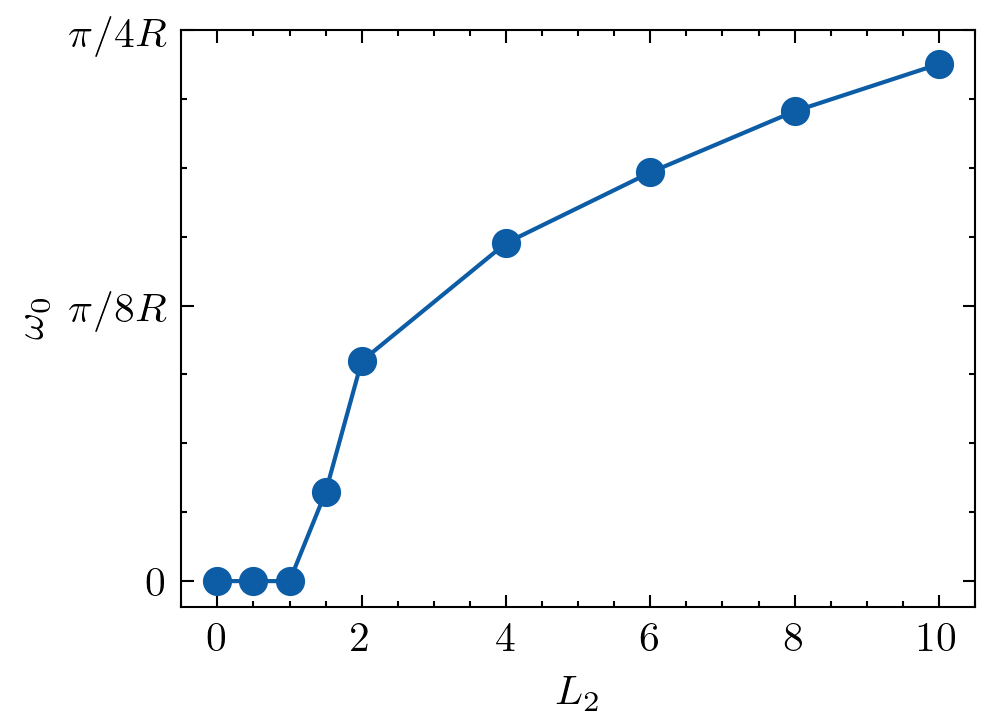}
        \caption{}
        \label{fig:equilibrium-twists}
    \end{subfigure}

    \caption{(a), Minimum free energy per unit length $F$ plotted against $\omega$ for three values of the anisotropic elastic constant $L_2$.
    The curves shown are parabolic fits to the data to estimate the location of the minimum, $\omega_{0}$, marked with the symbol $\times$ in the figure.
    (b), $\omega_{0}$  shown as a function of twist anisotropy $L_2$.
    Dimensional units are used, with $R$ being the radius of the capillary.}
    \label{fig:twisted-energy-fig}
\end{figure}

We have found that, once the configuration is initialized with a particular $\omega$, and using a long cylinder which is commensurate with $\pi/\omega$, iteration of Eq. \eqref{eq:Q-tensor-eom} does not result in any appreciable local change in coiling; rather the only disclination motion observed is a change of distance between the lines and the cylindrical axis. 
Calculations have also been conducted with Dirichlet boundary conditions on the cylinder caps, with the configuration fixed according to Eqs. \eqref{eq:Q-2D-definition}-\eqref{eq:rotation-matrix-definition}, yielding similar results. 
Therefore, in the analysis below, we set the length of the cylindrical cavity equal to half of the period corresponding to $\omega$ for $\omega \neq 0$, and we use a flat $2D$ configuration when analyzing uncoiled configurations. 
The resulting free energy landscape is then obtained as a function of $L_2$ and $\omega$ (Fig. \ref {fig:twisted-energy-fig}). 
Figure \ref{fig:all-L2-plot} shows representative curves of free energy per unit length versus wavenumber for several values of twist anisotropy. Each curve shows a distinct minimum away from $\omega = 0$, which increases with $L_2$.
Figure \ref{fig:equilibrium-twists} shows the value of the minimum wavenumber $\omega_0$ as a function of the twist anisotropy parameter $L_2$. Our results suggest a continuous bifurcation at $L_2 \approx 1.5$, below which the uncoiled configuration is the ground state, and above which the ground state coiling increases with increasing $L_2$ until a constant $\omega$ value is approached. We note that another branch, mirrored along the $L_2$ axis is implied, given that the free energy is invariant upon reflection over any plane.

Further insight into the nature of the ground states can be obtained by analyzing the director field in the vicinity of the disclination lines.
The disclination density tensor $D_{ij} = \epsilon_{i\mu\nu} \epsilon_{j\alpha\beta} \partial_{\alpha} Q_{\mu \delta} \partial_{\beta}Q_{\nu \delta}$ is computed
\cite{re:schimming22,re:schimming23}, and its dyadic decomposition near the line is used $\mathbf{D} = \frac{S_{N}^{2}}{a} \Omegahat \otimes \That $.
The unit vectors $\That$ and $\Omegahat$ are the local tangent to the disclination line, and the normal to the director rotation plane respectively.
In particular, $\That \cdot \Omegahat = \pm 1$ corresponds to a $\pm 1/2$ wedge disclination, while $\That \cdot \Omegahat = 0$ corresponds to a purely twist disclination.
\begin{figure}
    \centering
        \includegraphics[width=\columnwidth]{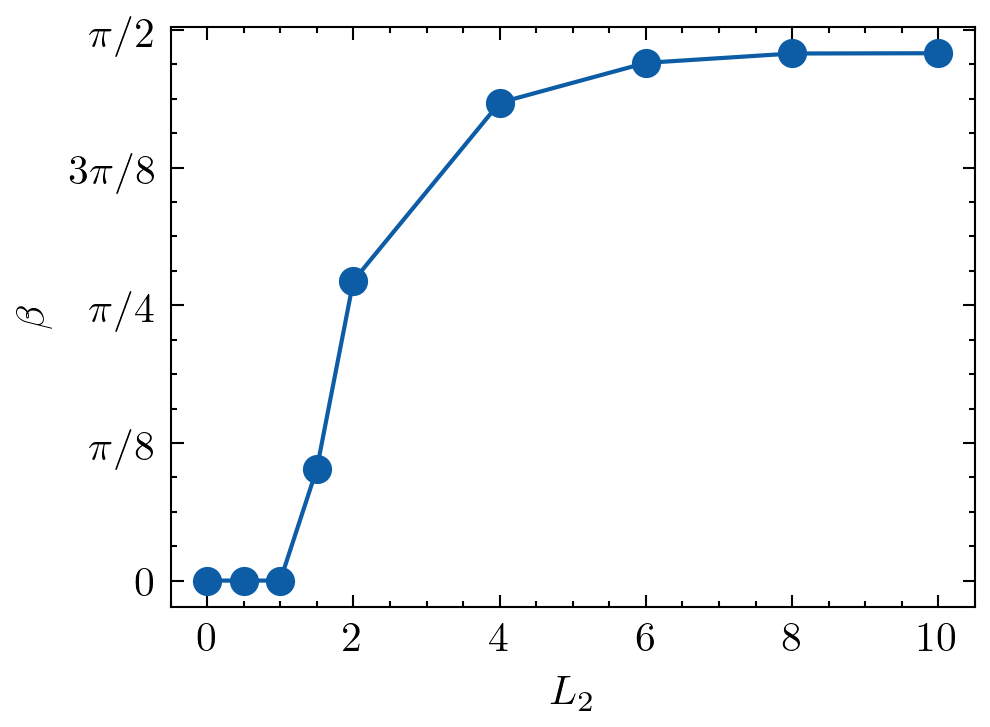}
    \caption{$\beta$, the angle between $\Omegahat$ and $\That$ plotted against twist anisotropy $L_2$.}
\label{fig:omega-T-angle}
  \end{figure}
Therefore the twist character of the disclination line is given by the angle $\beta = \arccos \left( \Omegahat \cdot \That \right)$, which for a doubly coiled configuration is constant along the line.
Figure \ref{fig:omega-T-angle} shows $\beta$ as a function of $L_2$.
The angle $\beta = 0$ for $0 \leq L_2 \lesssim 1.5$, and approaches $\pi / 2$ as $L_2$ increases, indicating that the director remains in-plane (wedge disclinations) when the disclinations are uncoiled, and the disclinations take on a twist character in coiled configurations.

A possible explanation for the energetic cause of the disclination coiling involves three competing factors: the geometry of the disclination lines (i.e. twist vs. wedge); adherence to homeotropic boundary conditions along the curved boundary of the domain; and the length of the disclination lines. For small twist elastic constant ($L_2 > 1.5$) it is energetically favorable for disclinations to take on a twist character, thereby lowering the amount of bend and splay in the configuration. However, this causes the director near the disclinations to point out of the cylindrical plane (Fig. \ref{fig:TP-x-y}). To maintain consistency with the homeotropic boundary conditions, which remain within the cylindrical plane, additional distortions are introduced between the disclinations and the boundary.
If the disclinations coil, however, the angle between the tangent vector of the disclination lines and the cylindrical plane becomes smaller than perpendicular. As a result, the director near the core may maintain a smaller angle with the cylindrical plane -- thereby lowering necessary distortions to remain consistent with the boundary conditions -- while also decreasing its angle with the vector tangent to the disclination -- thereby relieving bend and splay in favor of twist distortions. The coiling, however, increases the length of the disclinations which is energetically expensive.

This argument is consistent with Figs. \ref{fig:twisted-energy-fig} and \ref{fig:omega-T-angle}.
For an elastically isotropic system ($L_2 = 0$) there is no energetic benefit to relieving bend and splay in favor of twist distortion, so the disclination maintains its wedge character with $\beta = 0$. Since the director remains in the cylindrical plane, the disclinations do not need to coil in order to maintain consistency with the boundary conditions, so $\omega_0 = 0$. As $L_2$ increases it becomes increasingly energetically beneficial to adopt a twist-characteristic, and at $L_2 \approx 1.5$ the energetic penalties from the increased disclination length and boundary conditions are overcome.  $\beta$ then increases until it approaches $\pi/2$, at which point twist is maximized, and the disclinations coil to compensate for increased $\beta$ until it saturates. At this point, allowing the director to lie more in-plane would not compensate for the energy from increased disclination length, and so $\omega$ approaches a maximum.

\section{Twisted and untwisted escaped configurations} \label{sec:TER-ER}

When the radius of the capillary is sufficiently large, the ground state is an escaped configuration (without any defects), although it can also be chiral (Fig. \ref{fig:ER}). The relative stability of both untwisted (ER) and twisted (TER) escaped radial configurations in a cylinder with homeotropic boundary conditions was already given in Ref. \cite{jeong_chiral_2015}. Since these configurations do not contain defects, their stability was analyzed by minimizing the Frank free energy. It was found that the untwisted ER configuration has a smaller energy unless the twist elastic constant is sufficiently small $K_2 \lesssim 0.27K$, where $K = K_{1} = K_{3}$, the splay and bend elastic constants respectively. We reexamine this configuration here by using a $Q$-tensor model instead, in order to determine the transition line between escaped and polar configurations, as well as the  effect of bend and splay contrast on the relative stability of ER and TER configurations.

A two dimensional circular domain with uniaxial homeotropic boundary conditions is considered, although the tensor order parameter $\Q$ has five independent components to allow its eigenvectors to point in the third direction.
This setup is equivalent to a three dimensional system which is uniform in the third, $z$, direction. The initial configuration for the iteration of Eq. (\ref{eq:Q-tensor-eom}) is taken to be uniaxial with director $\nhat$ parameterized by angles $\alpha$ and $\beta$,
\begin{equation}
    \nhat
    =
    \cos \alpha \sin \beta \hat{\mathbf r}
    + \sin \alpha \sin \beta \hat{\boldsymbol \varphi}
    + \cos \beta \hat{\mathbf z}
\end{equation}
where $\alpha$ is the angle made between the $x$-$y$ projection of $\nhat$ and the radial vector $\hat{\mathbf r}$, and $\beta$ is the angle between $\nhat$ and $\hat{\mathbf z}$.
For an ER system, a minimizer for the Frank free energy under the condition of $K_1 = K_3$ is given by $\beta = 2 \arctan \left(\frac{r}{R}\right)$ with $R$ the radius of the circular domain, and $\alpha = 0$ \cite{re:meyer73}.
All untwisted escaped configurations are initialized this way.

\begin{figure}
    \centering
    \begin{subfigure}{\columnwidth}
        \includegraphics[width=\textwidth]{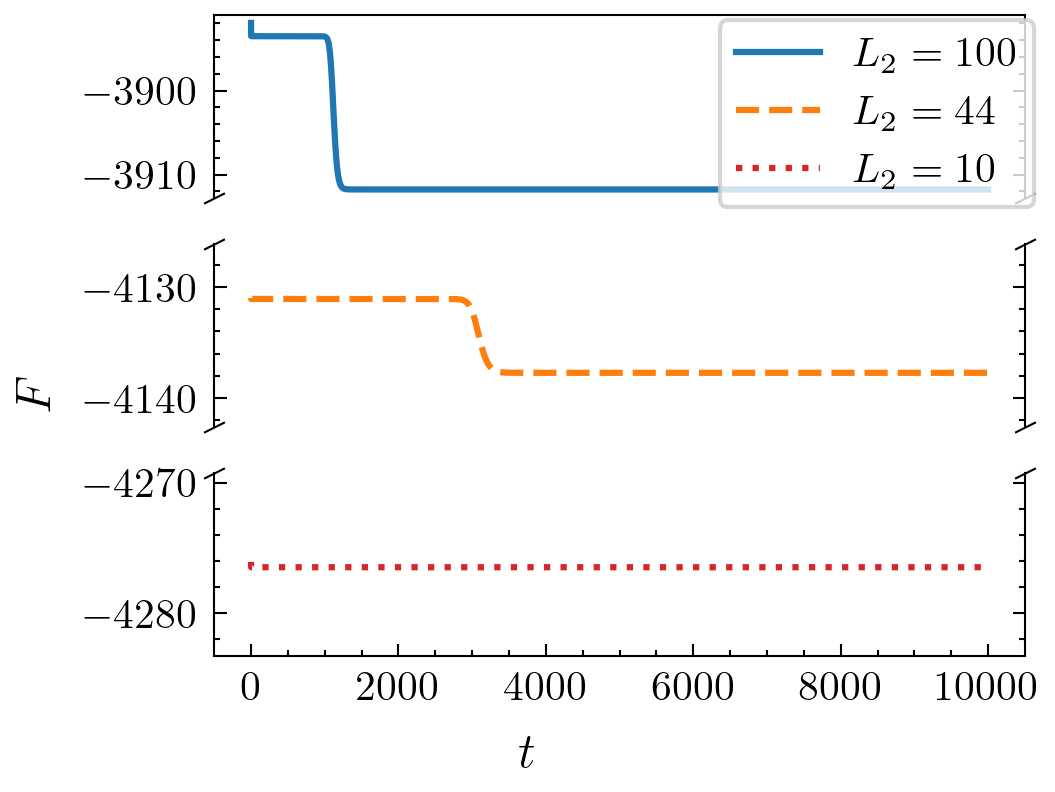}
        \caption{}
        \label{fig:ER-energy-examples}
    \end{subfigure}
    \hfill
    \caption{Free energy vs. time for ER-initialized configurations with twist elastic constant values $L_2 = 100, 44, 10$, and $L_3 = 0$ at $R = 100$.
    } \label{fig:ER-examples}
\end{figure}

To initialize all twisted escaped systems, $\beta$ is taken to be as in the untwisted case and $\alpha = \alpha_0 (1 - \frac{r}{R})$ for some $\alpha_0 \neq 0$ which characterizes the radial angle at the core. We take $\alpha_0 = 60^\circ$ based on the results of the director model analysis in Ref.  \cite{jeong_chiral_2015}. This value suffices to cause the system to decay to a twisted equilibrium configuration. To calculate minimum energy values, we iterate Eq. (\ref{eq:Q-tensor-eom})  until the relative change in free energy stabilizes.
For all configurations, $\left|dF/dt\right| / \left|F\right| < 5 \times 10^{-8}$.
Because of the metastability of the ER state (discussed below), a particular value for the relative change in free energy is not prescribed as criteria for stopping iteration in time.
Indeed, during the metastable energy plateaus in Fig. \ref{fig:ER-examples} the value of $dF/dt / F$ becomes less than $2 \times 10^{-14}$ so that in such cases configurations are allowed to continue to run until a further decay is observed, or a simulation time of $10,000 \, \tau$ has been reached.

To contextualize the discussion, it is helpful to consider the energy curves of three representative ER-initialized configurations as they decay (Fig. \ref{fig:ER-examples}). For all configurations, there is an immediate decay from the initialized configuration arising from the fact that the $Q$-tensor model ER energy minimizer is slightly different than the corresponding director configuration. The magnitude of this decay evidently increases as $L_2$ is increased.
For $L_2 = 100$ and $L_2 = 44$ there is an additional decay corresponding to the transition from ER to TER.
The magnitude of this decay increases with $L_2$, while the time at which the decay takes place decreases. 
For $L_2 = 10$, it is possible that there is a mechanism by which the ER configuration may dynamically decay to the TER configuration, but the numerical tolerances used here are not sensitive enough to allow such a decay.
This results from the fact that a Newton-Rhapson method with a finite tolerance must be used to iterate Eq. \eqref{eq:Q-tensor-eom} in time.
To get an idea of the precision involved, note that the finite element representation of $\partial \Q / \partial t$ has $L_\infty$ norm less than $10^{-10}$ before the algorithm is no longer able to iterate in time.
In the discussion that follows, ER-initialized configurations which are still in the metastable state (i.e. after the immediate decay, but before the decay into TER) are referred to as ER, and ER-initialized configurations which have decayed into a twisted state are referred to as ER $\to$ TER. TER-initialized configurations are simply referred to as TER.

\begin{figure}
    \centering
    \begin{subfigure}{\columnwidth}
        \includegraphics[width=\textwidth]{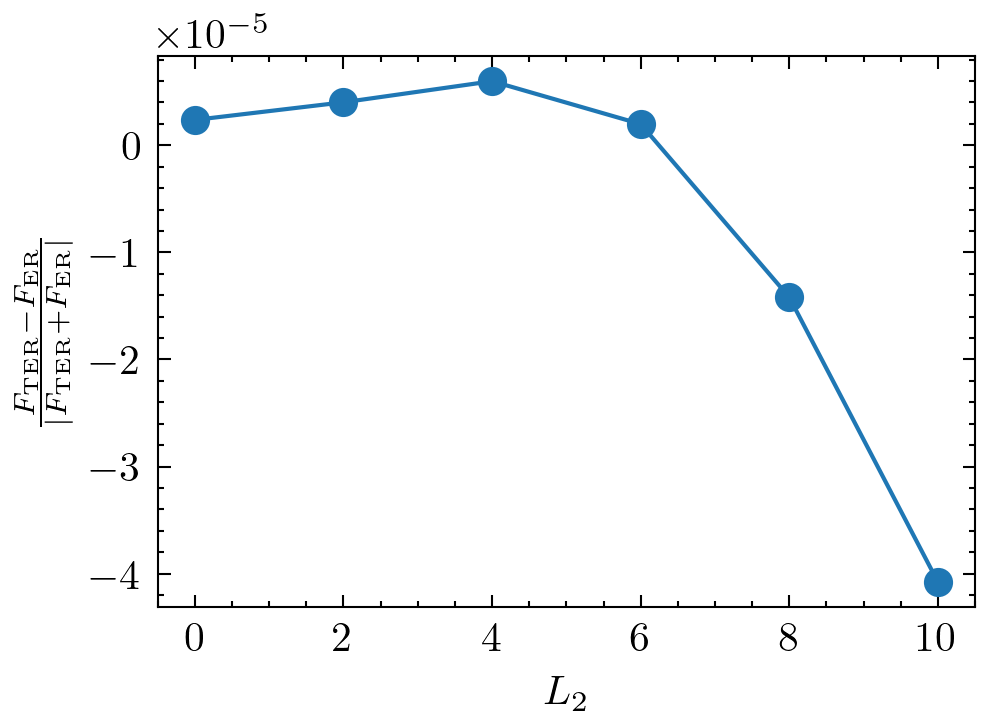}
        \caption{}
        \label{fig:ER-TER-energy-diff}
    \end{subfigure}
    \begin{subfigure}{\columnwidth}
        \includegraphics[width=\textwidth]{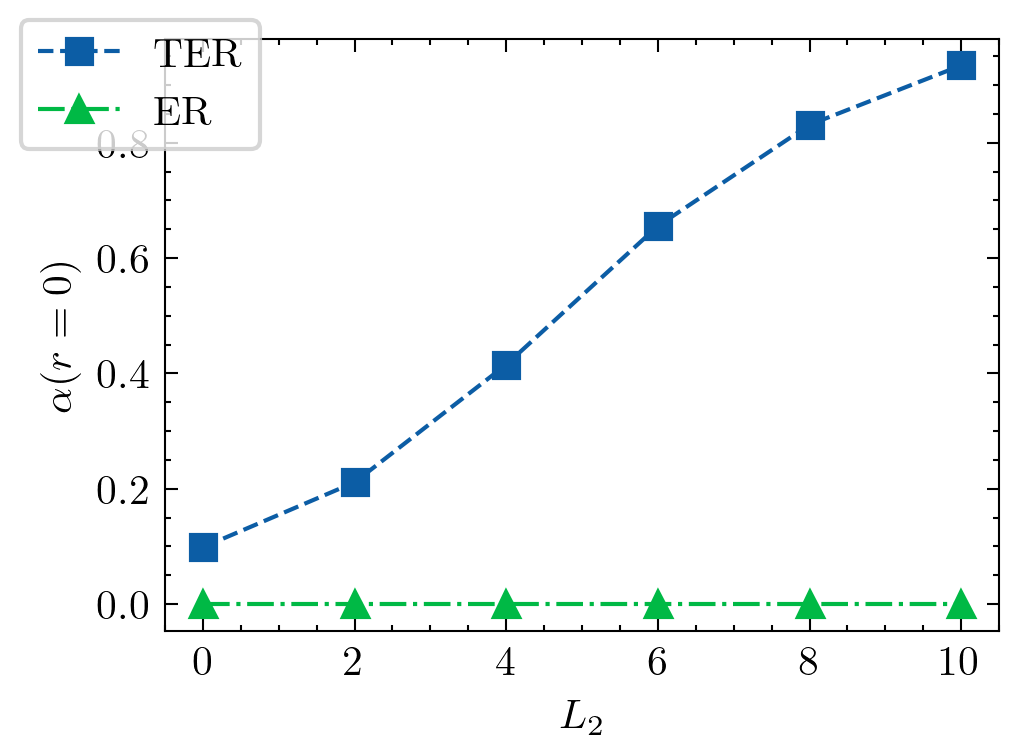}
        \caption{}
        \label{fig:alpha-angles}
    \end{subfigure}
    \hfill
    \caption{(a) Normalized energy differences between ER and TER configurations as a function of twist elastic constant $L_2$ evaluated for bend-splay elastic anisotropy constant $L_3 = 0$, and cylindrical radius $R = 100$.
    (b) $\alpha(r = 0)$ vs. twist elastic constant $L_2$ for relaxed ER and TER configurations with $L_3 = 0$ and $R = 0$.
    Here $\alpha$ is the angle made between the projection of the director into the cylindrical plane and the radial vector of the cylinder.
    } \label{fig:ER-TER}
\end{figure}

Free energy differences between twisted and untwisted configurations for $L_3 = 0$ and $R = 100$ are shown in Fig. \ref{fig:ER-TER-energy-diff}.
Both TER and ER configurations have almost identical free energies for a wide range of values of $L_2$, with the transition from untwisted to twisted ground state happening at $L_2 \approx 6.25$.
For the given parameters, the correspondence to the Frank free energy Eq. \eqref{eq:KL_mapping} gives the transition point at $K_2 \approx 0.24 \, K$, in rough agreement with \cite{re:jeong14}.
To characterize the amount of twist in the system, we note that $\alpha \to 0$ as $r \to R$ due to the strong homeotropic boundary conditions, and will approach a maximum as $r \to 0$ away from the boundaries. 
However, because the director escapes at the cylindrical axis, $\nhat = \hat{\mathbf z}$ so that $\alpha(r = 0)$ is technically undefined. 
In Fig. \ref{fig:alpha-angles}, $\alpha(r = 0)$ has been interpolated based on a parabolic fit of $\alpha(r)$ close to the cylindrical axis, and then plotted against $L_2$.
For $L_3 = 0$, when the configuration is initialized as ER, it remains untwisted ($\alpha(r = 0) = 0$) for subsequent iterations so long as it remains in the metastable state.
Therefore the ER configuration is a local free energy minimum.
Systems initialized in a TER configuration, $\alpha(r = 0) \neq 0$, remain twisted, with a twist angle that increases with $L_2$ (Fig. \ref{fig:alpha-angles}).
For $L_2 = 0$, $\alpha(r = 0)$ is very small which indicates that, in the isotropic limit, the TER initialized systems simply become an untwisted ER configuration.
This observation explains the near free energy degeneracy of both configurations in the range of small $L_2$ (and no bend/splay contrast).

\begin{figure}
    \centering
    \begin{subfigure}{\columnwidth}
        \includegraphics[width=\textwidth]{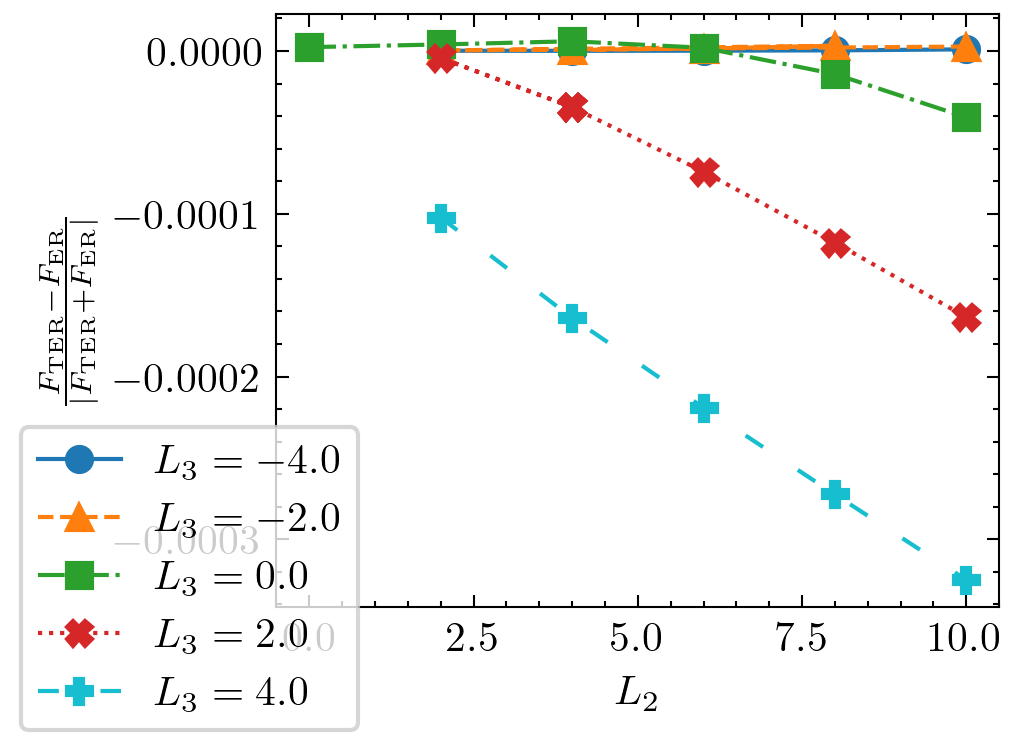}
    \end{subfigure}
    \hfill
    \caption{Normalized free energy differences in relaxed ER and TER configurations vs. twist anisotropy parameter $L_2$ for multiple values of bend-splay anisotropy parameter $L_3$.}
    \label{fig:ER-TER-energies-vs-L2}
\end{figure}

Figure \ref{fig:ER-TER-energies-vs-L2} shows the normalized free energy differences between twisted and untwisted configurations as a function of $L_2$ for multiple values of $L_3$. 
The free energy degeneracy is broken when $L_3 > 0$ which increases the bend elastic constant relative to both splay and twist elastic constants (Eqs. \eqref{eq:KL_mapping}).
When $L_{3} < 0$ the near degeneracy between ER and TER configurations persists to the largest values of $L_{2}$ that we have analyzed.
On the other hand, when $L_{3} > 0$, the TER configuration has the lowest free energy.

\begin{figure}
    \centering
    \begin{subfigure}{\columnwidth}
        \includegraphics[width=\textwidth]{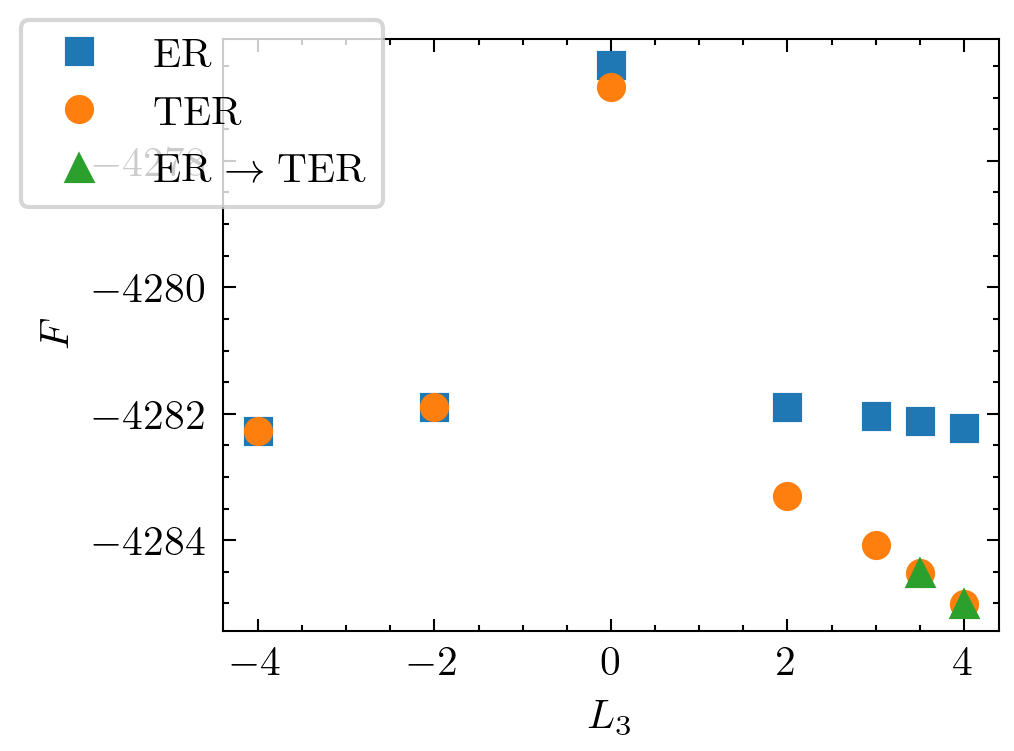}
        \caption{}
        \label{fig:ER-TER-energy-diff-anisotropic}
    \end{subfigure}
    \begin{subfigure}{\columnwidth}
        \includegraphics[width=\textwidth]{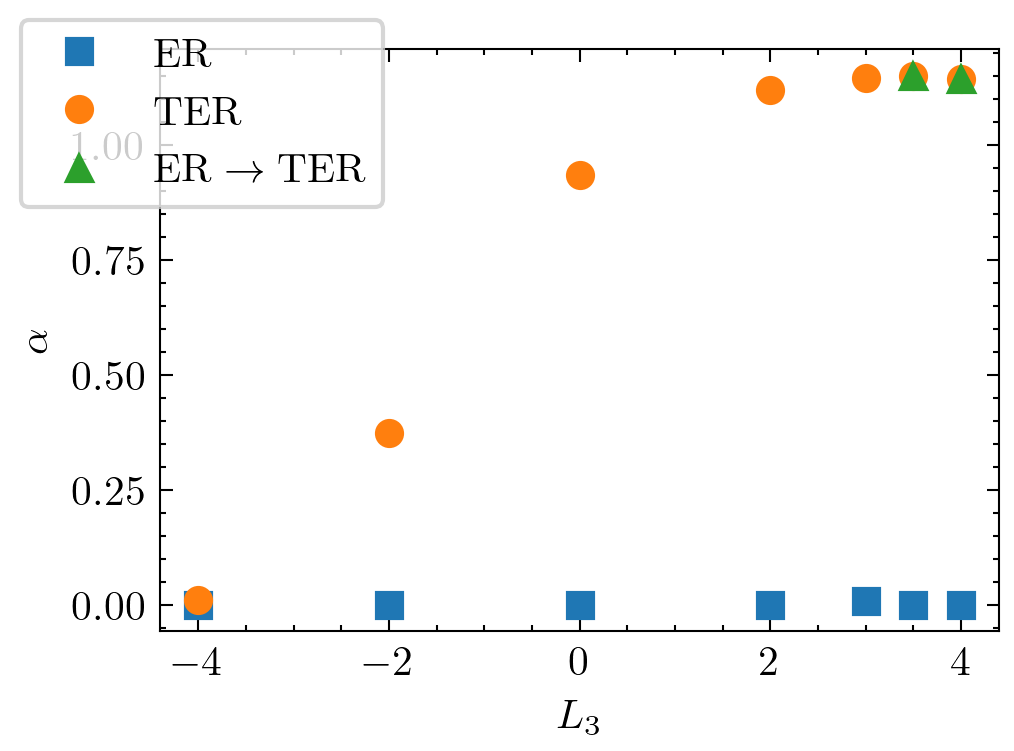}
        \caption{}
        \label{fig:alpha-angles-anisotropic}
    \end{subfigure}
    \hfill
    \caption{(a) Energy differences between ER and TER configurations as a function of elastic constant $L_3$ evaluated for twist elastic constant $L_2 = 10$ and cylindrical radius $R = 100$.
    (b) $\alpha(r = 0)$ vs. elastic constant $L_3$ for ER and TER configurations with $L_2 = 10$ and $R = 0$.
    For simulations in which there is an ER $\to$ TER decay as in Fig. \ref{fig:ER-examples}, both the metastable (ER) and final ground (ER $\to$ TER) configurations are shown.
    } \label{fig:ER-TER-anisotropic}
\end{figure}

Further analysis of the metastability of the two configurations is presented in Fig. \ref{fig:ER-TER-anisotropic}.
We set $L_2 = 10$ constant, vary $L_3$, and determine whether a configuration initialized as twisted or untwisted decays to the other after a long time of integration of Eq. \eqref{eq:Q-tensor-eom}.
The free energy at long times is also computed, and compared between the two configurations (apparent hysteresis).
When $L_3 < 0$, configurations initialized in the twisted TER configuration quickly decay to the untwisted ER configuration, whereas configurations initialized untwisted remain untwisted.
This is despite the fact that their free energies are very similar (Fig. \ref{fig:ER-TER-energy-diff}).
On the other hand, for large and positive values of $L_{3}$, the untwisted ER is seen to decay after a period of metastability to the twisted TER, which now becomes the lowest free energy state.
There is an intermediate range $0 < L_{3} \lesssim 3$ within which the ER remains stable during the time of integration studied, yet has higher free energy than the system initialized in a TER configuration.
Our results therefore suggest that the ER configuration is the ground state for $L_{3} < 0$ (splay elastic constant larger than bend), and the TER configuration for $L_{3} > 0$ (bend larger than splay), However, the ER configuration appears to remains metastable for a range of positive values of $L_{3}$ to the extent that we have not been able to observe its decay into a TER in the time span considered in our numerical calculations.

\section{Chiral state bifurcation diagram} \label{sec:phase-diagram}

Untwisted nematic liquid crystal configurations in a cylindrical cavity subjected to homeotropic boundary conditions have been compared to their twisted counterparts, in terms of relative energetic stability and configuration geometry.
In this section, we consider the relative stability of five such configurations studied thus far -- both twisted and untwisted -- over a range of capillary radii $R$ and twist anisotropy $L_2$ values.
This extends previous studies which have considered the relative stability of untwisted configurations for various temperatures and capillary radii, either under the assumption of elastic isotropy or for fixed $L_2 < 0$ \cite{shams_theoretical_2014, re:yan02}.
In order to allow for elastic anisotropy of the nematic and configurations comprising disclinations, the singular potential method has been used for the determination of the free energy of the configurations. 
When the twist elastic constant $K_2$ ($L_2$) is sufficiently small (large) as compared to bend or twist, ground states of broken chiral symmetry are found, both in defected and escaped configurations.

\begin{figure}
    \centering
        \includegraphics[width=\columnwidth]{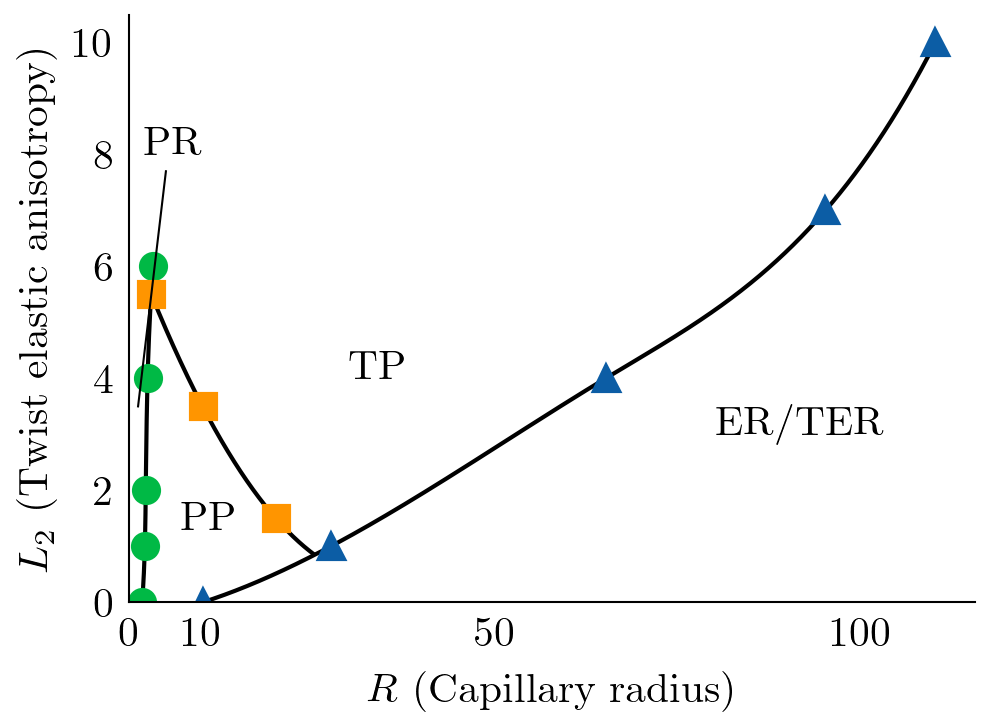}
    \caption{Regions of stability of the various configurations studied.
    Transitions based on energies of relaxed configurations are denoted with markers which are distinct for each transition, and lines are interpolated.}
     \label{fig:phase-diagram}
\end{figure}

Our results concerning the ground state are summarized in  Fig. \ref{fig:phase-diagram} in terms of the radius of the cavity $R$ and the twist anisotropy constant $L_{2}$. 
Five different configurations are shown, three that are achiral: a polar radial (PR) featuring a single disclination of charge +1 along the axis of the cylinder with director in the cylinder plane (Fig. \ref{fig:PR-diagram}, \ref{fig:PR-x-y}); a polar planar (PP) consisting of a pair of straight, parallel disclinations each of charge +1/2, also parallel to the cylinder axis (Fig. \ref{fig:PP-diagram}, \ref{fig:PP-x-y}); and the escaped radial (ER) configuration, which is defect free (Fig. \ref{fig:ER-x-y}, \ref{fig:ER-x-z}). 
The other two are the chiral counterparts: the twisted polar (TP) configuration in which two +1/2 disclination lines coil around each other forming a double helix configuration (Fig. \ref{fig:TP-diagram}, \ref{fig:TP-x-y}); and the twisted escaped radial configuration (TER) which remains defect free, but exhibits twist along the center of the capillary (Fig. \ref{fig:TER-x-y}, \ref{fig:TER-x-z}).
In the isotropic limit of $L_2 = 0$, the achiral PR, PP, and ER configurations are the ground states, in order of increasing capillary radius $R$. 
For capillary radii on the order of the disclination size, the achiral PR configuration continues to be the ground state. 
However, as $L_2$ is increased the range of $R$-values over which the single-defected PR is the ground state increases slightly.
This is consistent with previous studies in which the PR-PP transition happens at a larger $R$ value when temperature is increased \cite{shams_theoretical_2014, re:yan02}.
Because the PR and PP configurations contain no twist distortion modes, the effect of increasing $L_2$ is purely to increase the strength of long-range elastic alignment relative to short-range molecular alignment, the latter mediated by the parameter $\kappa$.
Since $\kappa \sim 1/\sqrt{T}$, lowering the effect of the short-range molecular alignment term has the same effect as increasing the temperature, hence why the PR-PP transition line curves to the right as in previous studies.

For capillary radius larger than the disclination size, but not yet macroscopic, the two-defected PP and TP configurations dominate. 
The former is the ground state closer to the isotropic limit, while the latter (chiral) configuration becomes the ground state as $L_2$ ($K_2$) is increased (decreased). 
The critical value of $L_2$ at which this happens decreases as the capillary radius is increased.

For very large capillary radii, escaped configurations have the lowest free energy, with the achiral (ER) and chiral (TER) configurations having nearly degenerate energy in the region which was studied, as discussed in section \ref{sec:TER-ER}.  
As noted above, the ER configuration becomes the ground state in the limit of $L_2 \to 0$, while the TER configuration becomes the ground state as $L_2$ becomes very large. 
In this range, splay-bend anisotropy breaks the degeneracy when $L_{3} > 0$ (bend elastic constant larger than splay), but not in the opposite case of $L_{3} < 0$. 
In the former case, splay-bend anisotropy is found to favor twisted escaped configurations.

As discussed in section \ref{sec:TP}, the disclination coiling wavenumber $\omega$ of two-defected configurations is never observed to dynamically change, so that neither PP $\to$ TP nor TP $\to$ PP transitions are observed.
Both TER $\to$ ER and ER $\to$ TER are dynamically observed, though the latter is only numerically resolvable for large $L_2$, given the nearly degenerate energy otherwise.
As described in section \ref{sec:TER-ER}, the TER $\to$ ER transition is a smooth untwisting, while the ER $\to$ TER transition happens suddenly from the untwisted metastable state, to the highly twisted ground state.
No transition is ever observed to or from the TP state, and when a PR configuration is initialized in the TP region of the phase diagram with periodic boundary conditions and length commensurate with $\omega_0$ observed for the $(R, L_2)$ parameter set, the configuration decays into a PP state.
It is unclear why this PR configuration decays into the metastable PP state instead of the TP true ground state, and the question of how, dynamically, TP configurations arise in experiments remains open.

\section*{Conclusions}

We have presented an analysis of nematic configurations which exhibit broken chiral symmetry under cylindrical capillary confinement with homeotropic anchoring. For configurations consisting of a pair of coiled +1/2 disclinations forming a double-helix, it is argued from the Frank free energy that ground state configurations are disallowed from having the director confined to the plane.
Rather, the disclinations take on a twist character in which the director near the disclination core approaches parallel to the disclination tangent as the twist anisotropy parameter $L_2$ is increased.
A critical value of this parameter is found at which the ground state transitions from the straight, parallel disclination PP configuration to the coiled disclination TP configuration, which decreases as capillary radius $R$ is increased.

The escaped ER and TER configurations are also studied, and a transition is found at which the ground-state becomes twisted for large $L_2$, though their energies are nearly degenerate.
Introducing bend-splay anisotropy by increasing the $L_3$ parameter breaks this degeneracy, and an abrupt transition from the metastable ER to the ground state TER is observed during cases in which their energy difference is sufficiently large.
The geometric structure of these configurations is also studied, with the ER configurations remaining untwisted while they are in the metastable state, and the TER configurations taking on an increasingly twisted character as $L_2$ is increased.

Finally, a bifurcation diagram is presented at fixed $\kappa = 8.0$ (related to temperature) in terms of $L_2$ as a function of capillary radius $R$.
There are regions in which each of the PR, PP, TP, ER, and TER configurations are ground states, and chiral configurations become more stable as $L_2$ is increased. 
We note, additionally, that for large capillary radius, the PP and TP configurations remain metastable, despite the fact that the escaped configurations are the true ground state.
We speculate that this is the reason why the PP and TP configurations are experimentally observed over long time-scales.

\section*{Conflicts of interest}
There are no conflicts to declare.

\section*{Data availability}

Data for this article is available at Zenodo at \href{https://doi.org/10.5281/zenodo.14902798}{https://doi.org/10.5281/zenodo.14902798}.
Code used to generate this data is available at Zenodo at \href{https://doi.org/10.5281/zenodo.14872507}{https://doi.org/10.5281/zenodo.14872507}.
Figs. 4-10 were generated with scripts available at the aforementioned code repository in the \texttt{app/analysis/figures} folder.
Data for Fig. 5 was processed in Paraview using a programmable filter available in the code repository as \texttt{app/analysis/paraview/programmable\_filter\_rotation\_angle.py}.
Data for Figs. 7b and 9b was processed in Paraview using a programmable filter available in the code repository as \texttt{app/analysis/paraview/pf\_director\_radial\_angle.py}, and then interpolated using the \texttt{app/analysis/plotting/plot\_alpha\_angle.py} script.

\section*{Acknowledgments}
This research has been supported by the National Science Foundation under contract DMR-2223707. This work used Expanse at the San Diego Supercomputing Center through allocation PHY170021 from the Advanced Cyberinfrastructure Coordination Ecosystem: Services \& Support (ACCESS) program, which is supported by U.S. National Science Foundation grants 2138259, 2138286, 2138307, 2137603, and 2138296. This research is also supported by the Minnesota Supercomputing Institute of the University of Minnesota.



\balance


\bibliography{references} 
\bibliographystyle{rsc} 

\end{document}